\documentclass[11pt,a4paper]{article}
\usepackage{jcappub}
\usepackage{mathrsfs}
\usepackage{dcolumn}
\usepackage{bm}
\usepackage{color}
\usepackage{ulem}
\usepackage{multirow}
\providecommand{\beqa}{\begin{eqnarray}}
	 
	 \providecommand{\rm}{\mathrm}
	\providecommand{\eeqa}{\end{eqnarray}}

\newcommand{\beq}{\begin{equation}}
	\newcommand{\eeq}{\end{equation}}

\newcommand{\cs}{c_{{}_S}}
\def\T{{\scriptscriptstyle {\rm T}}}

\def\46{{\scriptscriptstyle {\rm 4-6}}}
\def\24{{\scriptscriptstyle {\rm 2-4}}}

\title{EFT Compatible PBHs:\\ Effective Spawning of the Seeds for Primordial Black Holes during Inflation}
\author[a]{Amjad Ashoorioon,}
\author[a]{Abasalt Rostami}
\author[a,b]{Javad T. Firouzjaee}
\affiliation[a]{\small School of Physics, Institute for Research in Fundamental Sciences (IPM),\\
	P.O. Box 19395-5531, Tehran, Iran}
\affiliation[b]{\small Department of Physics, K. N. Toosi University of Technology,  P.O. Box  15875-4416, Tehran, Iran}
\emailAdd{amjad@ipm.ir}
\emailAdd{aba-rostami@ipm.ir}
\emailAdd{firouzjaee@kntu.ac.ir}

\abstract{Most of the inflationary scenarios that try to explain the origin of Primordial Black Holes (PBHs) from the enhancements of the power spectrum to values of order one, at the relevant scales, run into clashes with the Effective Field Theory (EFT) criteria or fail to enhance the power spectrum to such large amplitudes. In this paper, we unravel a mechanism for enhancing the power spectrum during inflation that does not use the flattening of the potential or reduction of the sound speed of scalar perturbations. The mechanism is based on this observation in the formalism of Extended EFT of inflation (EEFToI) with the sixth order polynomial dispersion relation for scalar perturbations that if the quartic coefficient in the dispersion relation is negative and smaller than a certain threshold, the amplitude of the power spectrum is enhanced substantially. The instability mechanism must arrange to kick in at the scales of interest related to the mass of the PBHs one would like to produce, which can be ten(s) of solar mass PBHs, suitable for LIGO events, or $10^{-17}-10^{-13}$ solar mass PBHs, which can comprise the whole dark matter energy density. We argue that the strong coupling is avoided for the range of parameters that the mechanisms enhance the power spectrum to the required amount.}

\keywords{Extended Effective Field Theory, Inflation, Primordial Black Holes}
\subheader{IPM/P-2019/018}
\begin{document}
	\maketitle
	\section{Introduction}
	\label{section:introduction}
	The LIGO discovery of the gravitational waves, from the merging of the binary of black holes (BH) \cite{TheLIGOScientific:2016src,Abbott:2016nmj}, revitalized the interest in the PBHs as the dark matter candidates, especially in the mass range ${\rm few}M_{\odot} \lesssim M_{\rm bh}\lesssim {\rm few}\times 10 M_{\odot}$ \cite{Bird:2016dcv,Sasaki:2016jop,Clesse:2016vqa}.
It is possible that these black holes are primordial and are born before nucleosynthesis, hence constitute all or part of the dark matter component of the energy budget of the universe.

Several mechanisms have been put forward to explain the formation of such PBHs during the early universe cosmology. One example of such a mechanism is bubble collision \cite{Hawking:1982ga}. In this scenario, a first-order phase transition from a metastable false vacuum, which proceeds through nucleation of the bubbles of the true phase, can lead to the formation of PBHs, if an enough number of bubbles collide at a point. This can be shown to be happening if the nucleation rate per unit four-volume is substantial but not instantaneous. The mass of such PBHs is proportional to the horizon mass, $\sim M_{{}_P}^2/H$. In \cite{Crawford:1982yz}, assuming that the QCD phase transition is a first-order one, the mass of most PBHs produced during the phase transition is shown to be of the order of $10^{30}$g, which is about $10^{-3}$  solar mass. In a similar context, the formation of a vacuum bubble or spherical walls during inflation is shown to be able to produce black holes after the termination of inflation \cite{Deng:2017uwc,Deng:2016vzb,Ashoorioon:2020hln}. The fate of the resulting black hole would depend on the mass or radius of the ensued bubble or domain wall, which can lead to an ordinary black hole or a baby universe separated from the parent FRW universe by a wormhole. In another scenario, shrinkage of a small fraction of cosmic string loops by a factor $\frac{1}{G\mu}$ leads to the formation of black holes \cite{Hawking:1987bn,Polnarev:1988dh}. Such black holes emit $\gamma$-rays, which their lack of observation puts an upper bound $G\mu\lesssim 10^{-6}$.  Incidentally, the bound on cosmic string tension from the limit on stochastic gravitational-wave background \cite{Sanidas:2012ee,Ringeval:2017eww} from the European Pulsar Timing Array and the PLANCK 2013 data is at most only an order of magnitude tighter \cite{Ade:2013xla}. However, more recent modeling of the cosmic string with realistic loop distribution with string microstructures, such as cusps and kinks, set a tighter bound of $G\mu\lesssim 10^{-11}$  on the string tension \cite{Ringeval:2017eww}. In the context of cosmic strings, it has recently been shown that PBHs can also form from the collapse of a small segment of cosmic string in the neighborhood of a cusp \cite{Jenkins:2020ctp}. Such a model only accounts for a small fraction of dark matter energy density, though.  Another way to produce PBHs is from the domination of exotic superheavy particles in the context of Grand Unified Theories (GUTs) \cite{Khlopov:1980mg}. This effect has been used to put some constraints on the GUTs.

The collapse of large density fluctuations is another mechanism of the formation of PBHs \cite{Carr:1974nx}. In this scenario, horizon reentry of a mode with large fluctuating amplitude during the ensuing radiation domination era leads to domination of gravity pressure in the over-dense region and its collapse to the black hole. Since inflation is the prime candidate for the generation of primordial fluctuations from which large-scale structures arise, it may also produce fluctuations at smaller wavelengths. This will give us unparalleled information about fluctuations at smaller wavelengths than the CMB scales and hence the nature of inflation. On the other hand with lensing, pulsar timing and other astrophysical techniques \cite{Schutz:2016khr,Monroy-Rodriguez:2014ula,Gaggero:2016dpq,Yokoyama:1998xd,Carr:2020gox}, we can put very tight constraints on the initial energy density of PBHs $\beta'(M_{{}_{\rm PBH}})=\rho_{{}_{\rm PBH}}/\rho$ in various mass ranges. These constraints can be used to limit the primordial power spectra of curvature perturbations in the inflationary models. Another advantage of this approach is that it can produce PBHs with different masses, from the subsolar to the supersolar, depending on the horizon size when corresponding modes with large amplitude re-enter the horizon. This feature is contrary to the astrophysical black holes that should have a mass above $3 M_{\odot}$.

In the context of standard inflationary scenarios where quantum fluctuations start from the vacuum, the inflaton potential determines the amplitude of primordial perturbations and their scale dependence. Observational limits on the power spectrum of curvature perturbations can then substantially help us constraining the inflationary models. At the moment, the latest probes of the primordial power spectrum with CMB scales \cite{Akrami:2018odb} with large scale structure surveys reveal the shape of the power spectrum on the scales between $\sim 1~{\rm Mpc}\ -\ 3000~{\rm Mpc}$ to be an almost scale-invariant power-law form.  On the other hand, limits on the PBHs abundances with different masses can be applied over a wider interval of scales \cite{Carr:1974nx, Carr:1993aq} {\it i.e.} $10^{-23}~{\rm Mpc}\ -\ 100~{\rm Mpc}$. Although the constraints on the PBHs abundances are not very tight in some mass ranges, their wide range of masses, which depends on the wavelength of the density perturbations when it re-enters the horizon, could be used to limit the shape of the primordial power spectrum in a much vaster range of scales than what is probed by the CMB and LSS surveys. If the density power spectrum of perturbations had an exactly scale-invariant power spectrum of order $\mathcal{O}(10^{-9})$ which is the measured value at the largest CMB scales, the PBHs abundance would be $\beta'(M_{{}_{\rm PBH}}) \sim e^{-10^8}$, which is too small. It essentially means that PBHs from inflation do not exist. To have a substantial value for the mass fraction of PBHs, one has to depart from scale invariance in scales of interest for the PBH formation. Different effects and phenomena, which fall into two different classes of gravitational and evaporation constraints, limit the PBH mass fraction depending on their mass \cite{Josan:2009qn}. In some mass ranges (more specifically $10^{-20} M_{\odot}\lesssim M \lesssim 10^{-17} M_{\odot} $) the abundance of PBHs has to be  $\beta'(M_{{}_{\rm PBH}}) \lesssim 10^{-22}$. Assuming that the power spectrum could be approximated by the scale-invariant power-law form in the region of interest, this, in general, would constraint the amplitude of the power spectrum of the primordial fluctuations \cite{Cole:2017gle} to be $\mathcal{P}_{{}_S} \lesssim \mathcal{O}(10^{-1}-10^{-2})$. However, since the mass variance is related to the power spectrum via an integration (see section \ref{PBHformation}), in addition to the amplitude, the shape of the power spectrum is also important. For instance,  using numerical analysis of the gravitational collapse and peak theory, one can show that for a power spectrum with a top-hat profile, while the enhancement in the power spectrum should be of the order $\mathcal{O}(10^{-2})$, this value grows to the order of $\mathcal{O}(10^{-1})$ for a narrow peak profile if it is assumed that the whole cold dark matter budget of the universe comes from the PBHs \cite{Germani:2018jgr}.

It was realized that \cite{Green:1997sz,Kim:1996hr} PBHs abundance places an upper bound on the spectral index of power-law power spectrum of primordial curvature perturbations\footnote{These limits arise from either this assumption that $\Omega_{{}_{\rm PBH,0} }< 1$ or from the evaporation of PBHs.} as $n_{{}_S} \leq (1.23 - 1.25)$. At this time, from CMB observations we can accurately measure $n_{{}_S}$ and we know that $n_{{}_S}=0.9649\pm 0.0042$ \cite{Akrami:2018odb}. As stated above, a power-law power spectrum with such spectral index in all scales never allows for a significant number of PBHs. For generating $30M_{\odot}$ PBHs in scales of about $k_{30M_{\odot}}\approx 3\times 10^{5}~{\rm Mpc}^{-1}$, the spectral index should have been around $n_{{}_S}\approx 1.85$ at the CMB pivot scale. One may assume that the power spectrum has a red tilt on the CMB scales compatible with the Planck 2018 data, but outside the probed scale obtains a blue spectral index at the scales relevant to the formation of PBHs. One can readily verify that the spectral index in the blue part should be as large as $n_{{}_S}\simeq 2.925$ or $1.582 \lesssim n_{{}_S}\lesssim 1.685$, respectively, to be able to accommodate the formation of PBHs with ${\rm few}\times 10 M_{\odot}$ or the ones with $10^{-17}M_{\odot}\lesssim M \lesssim 10^{-13}M_{\odot}$. Scenarios of running mass inflation \cite{Stewart:1997wg} can in principle accommodate such power spectra. However, with only running of the scalar spectral index in the observationally allowed range, $\frac{d n_{{}_S}}{d \ln k}=-0.0066\pm 0.0070$, one can not realize such a power spectrum. One then has to resort to the running of the running to achieve large amplitudes for the power spectrum at the scales of interest \cite{Drees:2011hb,Drees:2011yz}. Besides, to avoid the overproduction of the PBHs at small scales, one has to consider running of the running \cite{Carr:2016drx}. A working model would then need three parameters.  At the moment, the final verdict is that it is difficult to use these kinds of models to produce $30 M_{\odot}$ PBHs, although they can be used to produced PBHs in the mass range $10^{-17}M_{\odot}\lesssim M \lesssim 10^{-13}M_{\odot}$ that can compose the whole dark matter density \cite{Sasaki:2018dmp}.
Another way to accommodate an enhancement in the power spectrum at the scales of interest for PBH formation is considering a near inflection point \cite{Allahverdi:2006iq} in the inflaton potential, {\it i.e. } $V'(\phi_{{}_0})\simeq V''(\phi_{{}_0})\simeq 0$ \cite{Garcia-Bellido:2017mdw,Germani:2017bcs}. The enhancement could be roughly understood noting the fact that the power spectrum is proportional to $V^{3/2}/V'$, and for the modes that exit during the inflection point inflation, the power spectrum should increase. The detailed numerical calculation verifies this enhancement, although the shape of the power spectrum will not be completely the same as what the slow-roll expression suggests. Of course, these models require careful fine-tuning near the inflection region to obtain the scalar spectral index and amplitude of the power spectrum at the CMB scales compatible with the Planck data. It has been shown that the maximum enhancement of the power spectrum which can be achieved through such scenarios is $\mathcal{P}_{{}_S}\sim 10^{-4}$ at most \cite{Motohashi:2017kbs}, which is not suitable for PBH production \cite{Germani:2018jgr}. There are still models that claim that they have obtained a larger amplitude for density perturbations through this mechanism \cite{Cicoli:2018asa} suitable for the sufficient production of PBHs.  Temporary slow-roll violation may also occur through double inflation models like hybrid inflation, where two scalar fields control inflationary dynamics \cite{Clesse:2015wea}. In the simple hybrid model, we have a massive scalar field with a false vacuum energy density and the other scalar field, which controls the mass of the former. Inflation ends at a critical value where the massive field becomes tachyonic \cite{Linde:1993cn}.  If, on the other hand, the other direction does not become too steep to end inflation, like how it does in the case of the simple hybrid model, the power spectrum will obtain a large amplitude at the scales around the transition \cite{Kawasaki:2006zv}.

One can also assume a periodic structure in the inflationary potential at the small scales that leads to PBH formation. When the inflaton passes through such structures, the parametric resonance gives rise to the large curvature perturbations and consequently the PBH formation upon reentry and production of stochastic gravitational waves \cite{Cai:2019bmk}.

Another parameter in the power spectrum that one can fiddle with to enhance it at the scales of interest is the sound speed. The power spectrum for scalar perturbations is inversely proportional to $\cs^2$, and hence by reducing it, one can achieve an enhancement in the power spectrum \cite{Ozsoy:2018flq}. The reduced sound speeds could be realized in inflationary models in which the inflaton's kinetic terms are non-canonical \cite{ArmendarizPicon:1999rj}, where an example of it in string theory is DBI inflation \cite{Alishahiha:2004eh}. However, in such scenarios, even if one assumes that the amplitude of the power spectrum is enhanced to the conservative value of $0.1$, $
\cs$ should be lowered to values as small as $0.0014$. This is smaller than the lower bound that the Effective Field Theory of Inflation  (EFToI) sets on $c_{{}_S}$ from unitarity, {\it i.e.} $c_{{}_S}\gtrsim 0.003$ \cite{Cheung:2007st}. With $\cs \simeq 0.003$, in the horizon mass range $10^{-17}M_{\odot}\lesssim M \lesssim 10^{-13}M_{\odot}$, one can at most achieve enhancements up to $\simeq 10^{-4}$,  which as argued by \cite{Germani:2018jgr}, is hardly enough to account for the whole dark matter.

Besides the form of the potential or the speed of propagation of scalar perturbations, the dispersion relation for the scalar perturbations can affect the amplitude of the power spectrum, too \cite{Martin:2000xs,Ashoorioon:2011eg}. In particular, it has been shown that, in this case, it is possible to obtain a large amplitude for scalar perturbations. The dispersion relation could be realized in the formalism of Extended Effective Field Theory of Inflation \cite{Ashoorioon:2018uey}, which somehow legitimizes the $k^6$ correction to the dispersion relation of the scalar perturbation when it is accompanied by smaller orders. If in such a scenario, the coefficient of the quartic term is negative and its value over the square of the coefficient of the sextic term is smaller than a threshold,  the power spectrum grows. Still, one has to make sure that the sixth order correction at the horizon crossing is subdominant relative to the other term, to be able to say that the effective field theory is not strongly coupled at horizon crossing. We will make sure that in the scenario that we discuss, for the numerical values of the coefficients that lead to copious PBH formation, this criterion is honored.

The structure of the paper is as follows. First, we review the formalism of EEFToI \cite{Ashoorioon:2018uey} which now has to incorporate time-dependence for the coefficients of the operators in the unitary gauge action of the EEFToI. In particular, the time dependence should provide us with a negative quartic coefficient in the dispersion relation, which should kick in about the scales of interest for PBH formation. We resort to numerics to find the scale-dependent power spectrum, which, in particular, shows a bump that can be arranged to be of the order of one by tuning the negative quartic coupling to be smaller than a threshold. We show that the width of the bump in the spectrum is the same as the time span for which the quartic coupling is negative, but it starts off with a delay with respect to the onset of the change in the dispersion relation. This means that it only leads to the boost of the spectrum for the modes that are still about three e-folds within the Hubble radius when the quartic coefficient becomes negative.
Similarly, the termination of the epoch of the negativity of the quartic coefficient in the dispersion relation will not be shutting down the amplification in the power spectrum immediately. We argue that the strong coupling is avoided for the parameters we have chosen to produce the PBHs with the desired mass range. We finally conclude the paper in the last section and lay down plans to continue this research line.

	\section{PBH from Extended EFT of Inflation}
	
	\subsection{Extended EFT of Inflation and Sixth Order Dispersion Relations}
	The main idea of Effective Field Theories (EFTs) is that the effect of the higher energy scales within a theory could be encoded in an infinite number of higher dimensional operators in such an approach. The idea has proven to be useful in quantifying the effect of high energy physics in particle physics and condensed matter and was extended to the area of inflation in \cite{Cheung:2007st,Weinberg:2008hq}. To establish the EFT of inflation for a single scalar field \cite{Cheung:2007st}, one should note that even though the space-time expands in de Sitter format during inflation, the accelerated expansion has finally come to an end. This implies the time-translation associated with the de-Sitter space-time has been spontaneously broken. Like any symmetry, its breaking suggests the existence of a Goldstone boson, which is adsorbed in the metric. In particular, in a single field inflationary background, one can go to a gauge in which the Goldstone boson, which transforms nonlinearly under the time diffeomorphism, is eaten by the metric that would acquire a longitudinal mode and thus will have three degrees of freedom. In this gauge, which is known as the unitary gauge, the inflaton fluctuations are absent{\it, i.e.} the surfaces of constant time and constant inflaton coincide. The remaining symmetries in this gauge are the spatial diffeomorphisms. One can write down the most general form of the action in the unitary gauge by including all the operators that respect these symmetries. The tree-level operators in the Lagrangian will be those that yield the single field slow-roll inflation. Higher mass dimensional operators would correspond to deviations from the slow-roll inflation. The approach would allow one to consider the realization of K-inflation and Ghost inflation within a single framework. However, the modification of the dispersion relation beyond the quartic order was not allowed by the original work. Based on the energy scaling of the Goldstone boson, $\pi$, they argued that when the dispersion relation $\omega^2\propto k^{2n}$ with $n\geq 3$, the interacting operators become dominant at the infrared and invalidate the effective field theory approach. However, as discussed in  \cite{Ashoorioon:2018uey, Ashoorioon:2018ocr}, this is not necessarily the case if the unitary gauge action renders lower-order corrections to the dispersion relation dominant at the horizon crossing. This is what motivated the formalism of the Extended Field Theory of Inflation (EEFToI). Whether the formalism allows going beyond the sixth order dispersion relation is something that one has to investigate thoroughly, but the possibility of such a construction will not bear upon the main theme of this paper. What is important to our proposal here is that there is an intermediate phase of domination by a term in the dispersion relation that comes with a negative sign. This intermediate phase is encompassed between at least two phases in which the dispersion relation is positive definite. In the deep UV, the positive definiteness of the dispersion relation helps us stabilize the fluctuations and define a unique positive frequency WKB vacuum. As long as the modes exit the horizon in the regime of domination of quadratic or quartic part of the dispersion relation, one can show that the Effective Field Theory is still viable. On the other hand, if the mode spends long enough under the influence of the term that comes with a negative contribution in the dispersion relation, the amplitude of fluctuations can get amplified while their wavelengths are still smaller than the Hubble length.  Hence, beyond the sixth order corrections to the dispersion relation is irrelevant to our proposal in this paper.

As noted above, once one goes to the unitary gauge, the most general action up to mass dimension four that can be written from the operators that respect the remaining spatial diffeomorphism, take the form,
\beqa \label{EEFToI}
\mathcal{L}
&=&
M_{\rm Pl}^2
\left[
\frac12\, R
+\dot H\,  g^{00}
-\left(3\, H^2 +\dot H\right)
\right]
\nonumber\\
&+& \frac{M_2^4}{2!}\,(g^{00}+1)^2
+\frac{\bar M_1^3} {2}(g^{00}+1) \delta K^\mu_{\ \mu} -\frac{\bar M_2^2}{2}\, (\delta K^\mu_{\ \mu})^2
-\frac{\bar M_3^2}{2}\, \delta K^\mu_{\ \nu}\,\delta K^\nu_{\ \mu}\nonumber\\
&+&\frac{\bar{M}_4}{2} \nabla^\mu g^{00}\nabla^\nu\delta K_{\mu \nu}
-\frac{\delta_2}{2}\,(\nabla_{\mu} \delta K^\nu_{\ \nu})^2
\nonumber\\
&-&\frac{\delta_3}{2} \,(\nabla_{\mu} \delta K^\mu_{\ \nu})(\nabla_{\gamma} \delta K^{\gamma\nu})-\frac{\delta_4}{2} \,\nabla^ {\mu}\delta K_{\nu\mu}\nabla^ {\nu}\delta K_{\sigma}^{\sigma}
\,,
\eeqa
where terms proportional to $(\nabla_{\mu} \delta K^{\nu\gamma})(\nabla^ {\mu} \delta K_{\nu\gamma})$ and $(\nabla_{\mu} \delta K^\mu_{\ \nu})(\nabla_{\gamma} \delta K^{\gamma\nu})$ are dropped as they lead to Ostrogradski ghosts \cite{Ostrogradsky:1850fid} in the second order action for tensor and scalar perturbations \cite{Ashoorioon:2018uey}\footnote{The only term that modifies the action of the tensor perturbations is $-\frac{\bar M_3^2}{2}\, \delta K^\mu_{\ \nu}\,\delta K^\nu_{\ \mu}$, which modifies the speed of propagation of tensor modes as,
	\beq\label{M3bar-negative-sign}
	c_\T^2=\left(1-\frac{\bar M_3^2}{M_{\rm Pl}^2}\right)^{-1}
	\ .
	\eeq.}. All the couplings in the unitary gauge action \eqref{EEFToI}, regardless of their mass dimension, can be functions of time, as any function of time respects the remaining spatial diffeomorphism symmetry too. We will use this time dependence later to create a time-dependence in the coefficients of the dispersion relation. Using the St\"uckelberg method to make the Goldstone boson $\pi$ explicit in the action and in the limit  where $\dot{H}\to 0$, one can find the following equation of motion for $u_k\equiv a\,\pi_k$, which is related to the gauge perturbations of the comoving hypersurface $\zeta_k=-\frac{H}{a}\, u_k$,
\begin{equation}\label{eom}
	\frac{d^2 u_k}{d x^2}+\left(c_{{}_S}^2 +\alpha x^2+\beta x^4-\frac{2}{x^2}\right) u_k=0\,.
\end{equation}
where $x\equiv k\tau$ and $u_k=a\,\pi_k$.  Above we have chosen $\delta_3=-3\delta_4$, to get rid of the the friction term in the equation of motion and also  mixing with gravity at super-Planckian momenta, when the mode is deep inside the horizon, ({\it c.f.} \cite{Ashoorioon:2018ocr}). The parameters in the equation of motion \eqref{eom} is defined as
where,
\beqa\label{disp-coeff}
c_{{}_S}^2&\equiv& \frac{F_2}{G_1}\,,\\
\alpha &\equiv& \frac{D_2}{G_1}\,,\\
\beta &\equiv& \frac{C_2}{G_1}\,,
\eeqa
where
\beqa\label{var2}
F_2&\equiv& -2 M_{\rm Pl}^2 \dot{H}-\bar M_1^3 H-3H^2 \bar M_2^2-\bar M_3^2 H^2-\frac{9}{2}  \delta_4 H^4-\bar{M}_4 H^3\,,\\
G_1&\equiv&-2 M_{\rm Pl}^2 \dot{H}+4 M_2^4-6\bar M_1^3 H-9 H^2 \bar M_2^2-3 H^2 \bar M_3^2-\frac{81}{2} H^4 \delta_4-9\bar M_4 H^3\,\\
D_2&\equiv& \bar M_2^2 H^2+\bar M_3^2 H^2-\frac{17}{2} \delta_4 H^4-\bar M_4 H^3\,,\\
C_2&\equiv& -2\delta_4 H^4\,.
\eeqa
As noted in \cite{Ashoorioon:2018ocr}, it is always possible to make $c_{{}_S}^2$ remain arbitrarily close to $1$. For this to be satisfied, the following relation between the couplings should be satisfied
\beq
4 M_2^4=5 \bar{M}_1^3 H+6 H^2 \bar{M}_2^2+2 H^2 \bar{M}_3^2+36 H^4 \delta_4+8 \bar{M}_4 H^3\,.
\eeq
\begin{figure}
	\centering{
		\includegraphics[width=9.5cm, height=7cm]{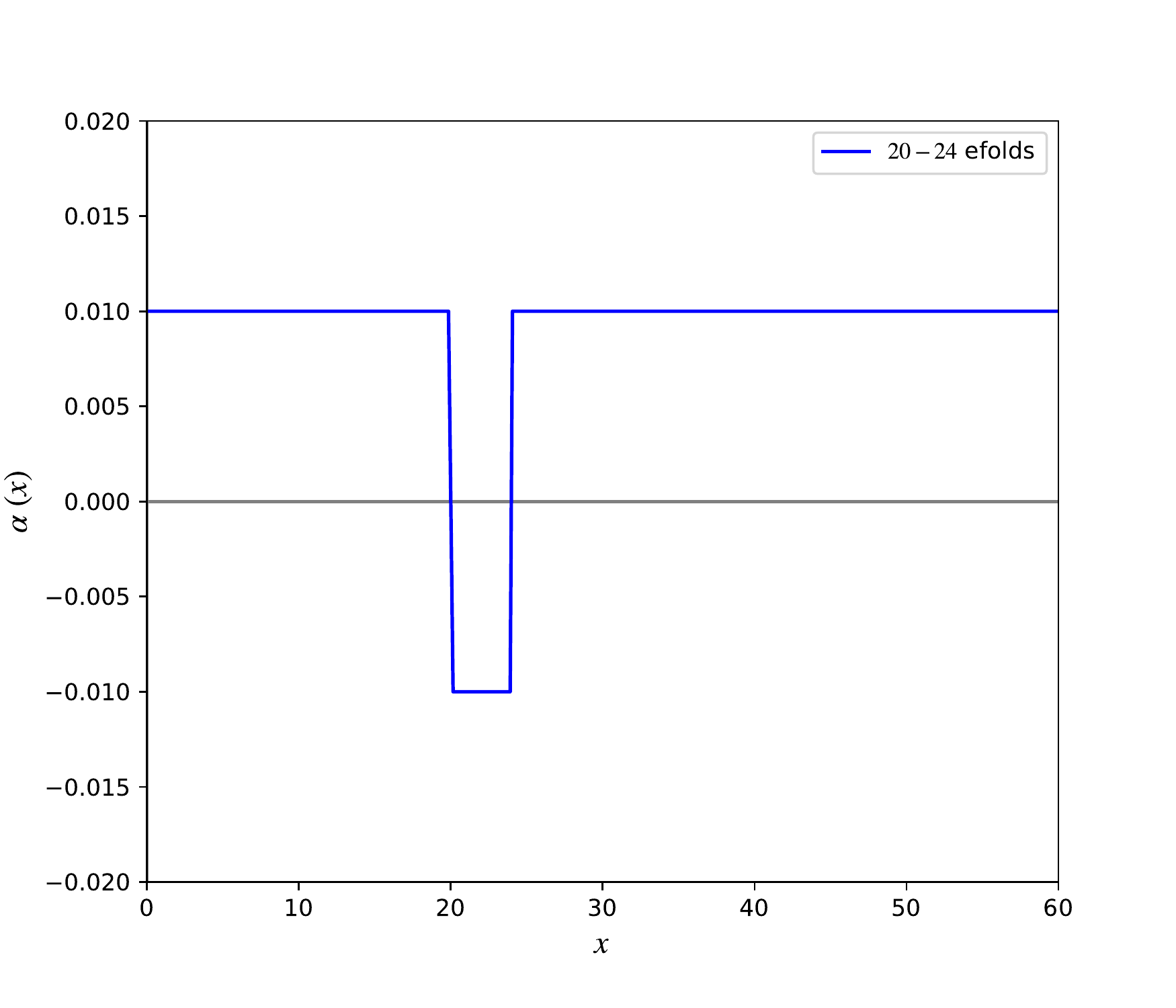}
		\caption{The plot shows the time-dependence of $\alpha$ with parameters $N_1=17$, $N_2=24$ and $\alpha_0=0.01$. $\Delta N$ is assumed to be $0.01$.}
		\label{fig:alpha}}
\end{figure}
This will help us in avoiding the strong coupling in the EEFToI, despite the large amplitude of the two-point functions possible. We also assume that $\beta>0$ to be able to define a positive frequency WKB-like vacuum. That means that the parameters $C_2$ and $G_2$ will have the same sign throughout. We will also assume that $\beta$ is positive throughout the inflation. For simplicity, we also assume that $\beta$ is almost constant and time-independent for the scales that exit during the last 60 e-folds. On the other hand, we assume that the parameter $D_2$ on the CMB scales have the same sign as $G_1$, but the time-dependence of the relevant couplings in the definition of $D_2$ is such that it flips sign in some intermediate-scales relevant for PBH formation. This could mean that the values of two of those parameters vary at two separate times of (phenomenological) interest almost instantaneously. As we will see in the next section, if the resulted quartic and sextic couplings satisfy certain relations, the amplitude of the fluctuations will increase substantially to the level that can lead to the formation of PBHs upon reentry during the ensuing radiation dominated phase. For simplicity, we assume that the absolute value of $\alpha$ before and after the sign flipping is the same. Such behavior in $\alpha$ can be approximated by the following function
\beq\label{alphat}
\alpha(t)=\alpha_0\left[1+\tanh\left(\frac{\exp(N_2)-\exp(t)}{\Delta N}\right)-\tanh\left(\frac{\exp(N_1)-\exp(t)}{\Delta N}\right)\right]
\eeq
where $\alpha_0>0$ and $t$ quantifies the number of e-foldings from the CMB scales to the end of inflation, where the end of inflation corresponds to $t=60$. The value of $\alpha$ swings from $\alpha_0$ to $-\alpha_0$ between $N_1<N<N_2$ e-foldings. The transition takes $\Delta N$ e-foldings, which we will assume to be much less than an e-foldings. A plot of $\alpha(t)$ as a function of number of e-folds is shown in Fig. \ref{fig:alpha}. As explained above, the values of $\cs^2$ and $\beta$ are assumed to remain positive and constant throughout this transition. In particular we assume that $\cs^2=1$ and $\frac{\beta}{\alpha_0^2}={\rm const.}$
The mechanism that we would like to suggest for the formation of PBHs is based on the observation in \cite{Ashoorioon:2018uey} is that for $\alpha<0$ the amplitude of scalar power spectrum grows, while the mode is still inside the horizon. In particular it was shown that if $\frac{\beta}{\alpha^2}<\frac{1}{4}$, the enhancement becomes large. Such fluctuations with large amplitude can lead to copious production of PBHs, once they re-enter the horizon during the radiation-dominated phase. The instability has to kick in at the right scales in order for the mass of the formed PBHs to fall in the phenomenologically motivated range. In particular, as we will see, there is a time delay in the rise and fall of the spectrum, which should be taken into account when the fundamental physics that has led to the growth of fluctuations is linked to the mass of the produced PBHs.
\subsection{Numerical Behavior of EEFToI Dispersion Relation}
In this section, we numerically solve the eq. (\ref{eom}) with the time-dependent expression, \eqref{alphat} and compute the power spectrum of the scalar perturbations for the modes that exit the horizon during the last 60 e-foldings on inflation. To achieve this goal, one must note that due to the extra quadratic and quartic terms in (\ref{eom}),  the standard Bunch-Davies vacuum mode can no longer be used as the initial condition for  the mode equation  $x\rightarrow -\infty$. In this case, since $\beta x^4$ is the dominant term in the far remote past, the normalized WKB positive frequency mode function, which  would minimize the Hamiltonian \cite{Martin:2000xs}
\begin{align*}
v_k(x) = \lim_{x\rightarrow -\infty} u_k(x)=i\, \sqrt{\frac{\pi x}{12}}\, H^{(1)}_{\frac{1}{6}} \left(x^3\frac{\sqrt{\beta}}{3}\right)
\end{align*}

Therefore, to solve equation (\ref{eom}) numerically, we will impose the following boundary conditions when $x\rightarrow -\infty$ \cite{Ashoorioon:2017toq}
\begin{eqnarray*}
	u_k(x) &=& v_k(x)\,, \\
	u'_k(x) &=& v'_k(x)\,. \\
\end{eqnarray*}

 It is useful to define the ratio $\gamma$ as follows,
\begin{equation}
	\gamma\equiv \frac{\mathcal{P}^{{}^{\rm EEFToI}}_{{}_S}}{\mathcal{P}^{{}^{\rm BD}}_{{}_S}}\,,
\end{equation}
where $\mathcal{P}^{{}^{\rm EEFToI}}_{{}_S}$ is the power spectrum of scalar perturbations in the EEFToI and $\mathcal{P}^{{}^{\rm BD}}_{{}_S}=H^2/{8\pi^2c_{{}_S}\epsilon}$ is the standard Bunch-Davis power spectrum of scalar perturbations. Due to the sextic and quartic correction terms in the dispersion relation of EEFToI, we expect that $\gamma$ deviates from unity. In \cite{Ashoorioon:2018uey} numerically solving for the \eqref{eom}, the values of $\gamma$ for $0.01\leq |\alpha| \leq 2$ and $0\le \beta \leq 2$ were obtained. The exploration revealed that for the positive values of $\alpha$ not only there is no cue of an enhancement in the power spectrum but, conversely, the factor $\gamma$ would get suppressed further as $\alpha$ and $\beta$ increases. For example for values of $\alpha=\beta=0.2$, one would get $\gamma=0.717$, and for $\alpha=\beta=1$ one obtains $\gamma=0.40$.
On the other hand, in the case where $\alpha < 0$, depending on the value of ratio $\beta/\alpha^2$ and $\alpha$, one could find considerable enhancements for the modulation factor, $\gamma$. In \cite{Ashoorioon:2017toq}, since the dispersion relation was motivated from its counterpart in the Minkowski background, the ratio was varied in the interval $\frac{1}{4}\leq \frac{\beta}{\alpha^2} < \frac{1}{3}$ to preclude the Minkowski counterpart of the dispersion relation becoming tachyonic. Nonetheless, even in this interval, one would get an enhancements as large as $\gamma\simeq 14.77$, for  $\alpha\simeq -0.2$ and $\frac{\beta}{\alpha^2}=\frac{1}{4}$. As stated above, in \cite{Ashoorioon:2018uey} however, the sixth order dispersion relation was motivated within the formalism of EFT of inflation, and {\it a priori}, there is no need to constrain the parameters such that $\frac{\beta}{\alpha^2}\geq \frac{1}{4}$. One is allowed to explore the smaller ratios so long as upon the horizon crossing, the sixth order correction to the dispersion relation is subdominant with respect to the quadratic or quartic terms.  For example, in \cite{Ashoorioon:2018uey}, it was noticed that for $\alpha=-0.01$ and $\beta/\alpha^2=0.2$, values as large as  $\sim 1000~\mathcal{P}^{{}^{BD}}_{{}_S}$ is obtained.
\begin{figure}
	\centering{
		\includegraphics[width=9.5cm, height=7cm]{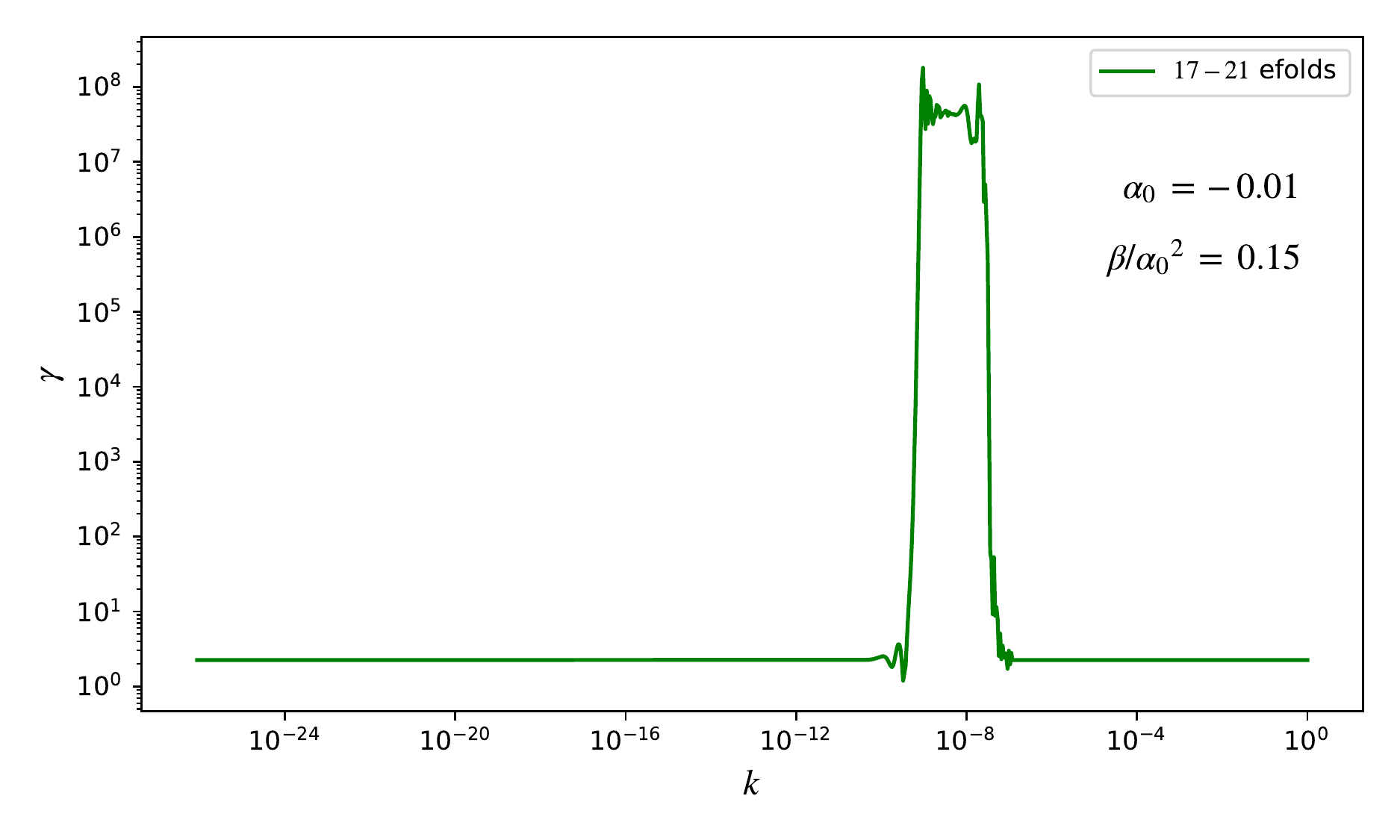}
		\caption{The figure in the right shows the resulting top-hat plot of $\gamma$ against wavenumber $k$. The instability appeared in scales $k=e^{-21}H-e^{-17}H$ would be corresponded to the PBHs formation of masses $10^{-17}M_{\odot}-10^{-13}M_{\odot}$ which can account for $10-100\%$ of the dark matter density.}
		\label{fig:alpha43}}
\end{figure}
This was suggestive that by reducing the ratio $\beta/\alpha_0^2$ further, one would be able to reach larger enhancements in the amplitude of scalar perturbations. Indeed with the ratio $\beta/\alpha_0^2\simeq 0.15$, the amplitude of the power spectrum could reach enhancements of order $10^8-10^9$, which would raise the amplitude of perturbations close to one. However, this should only happen for the scales of interest that lead to PBHs of proper mass. The possibility that in the EEFToI, the couplings in the unitary gauge action can be functions of time might properly accommodate this feature. In particular, it can be arranged that the dispersion relation leads to the standard Bunch-Davis power spectrum at the CMB scales accompanied by an enhancement of order $10^8-10^9 \mathcal{P}^{{}^{BD}}_{{}_S}$ for the scales of interest for PBH formation. As we stated above, for having a large value of $\gamma$, one has to impose $\beta/\alpha^2 < 0.25$  and $\alpha < 0$. We choose $\alpha$ a function of time which has  $\alpha_0>0$ before $N_1$ e-folds before the end of inflation and then plummets very fast to $-\alpha_0$ from $N_1$ to $N_2$ e-folds before the end of inflation. This can be achieved if one of the contributing factors in the definition of $\alpha$ has some sort of jump or phase transition in its evolution. Finally, $\alpha$ raises back to a small positive value, $\alpha_0$ due to transition in another contributing coupling in the unitary gauge action. The parameter $\alpha$ keeps this value until inflation ends (See Fig. \ref{fig:alpha}). As stated before, we set $\beta$ to  be constant such that $z\equiv\beta/\alpha_0^2<\frac{1}{4}$. A precise numerical analysis shows that for this choice of the coefficients of quartic and sextic corrections in the dispersion relation, we would see the general behavior anticipated for the modulation factor $\gamma(k)$, {\it i.e.} the ratio keeps being a positive constant value of order one during the wavenumber interval and then it reaches the desired large-amplitude once the transition to negative values happen. However, it is also noticed that the power spectrum does not momentarily reach large values once the instability happens. For the modes that are on the verge of horizon crossing, the power spectrum amplitude remains almost intact when the transition to negative values of quartic coupling occurs. It takes roughly about $N_d \sim 3$ e-folds until the mode starts feeling the dominance of the negative quartic term in the dispersion relation. Then for the total number of e-folds that the quartic coupling is negative, the power spectrum receives a large modulation factor. This would mean that the effect of quartic coupling being negative extends to the modes that exit the horizon when the quartic coupling is positive. This shows that the process of amplification of the mode occurs while the mode is inside the horizon. It is not something that occurs right at the horizon crossing. This point must be considered when ascribing the mass of produced PBHs to the onset and duration of the negative quartic dispersion relation, which comes from a more fundamental physics. The range of enhancement observed in the $k$ space would be
\beq
e^{-(N_1+N_d)} H\lesssim k\lesssim e^{-(N_2+ N_d)} H \,.
\eeq
Of course, the transition from $\gamma\simeq \mathcal{O}(10^8-10^9)$ to $\gamma \simeq \mathcal{O}(1)$ and vice versa do not occur smoothly. In between, there will be some modulated oscillations in the amplitude of the power spectrum.
\begin{figure}
	\centering{
		\includegraphics[width=0.7 \columnwidth]{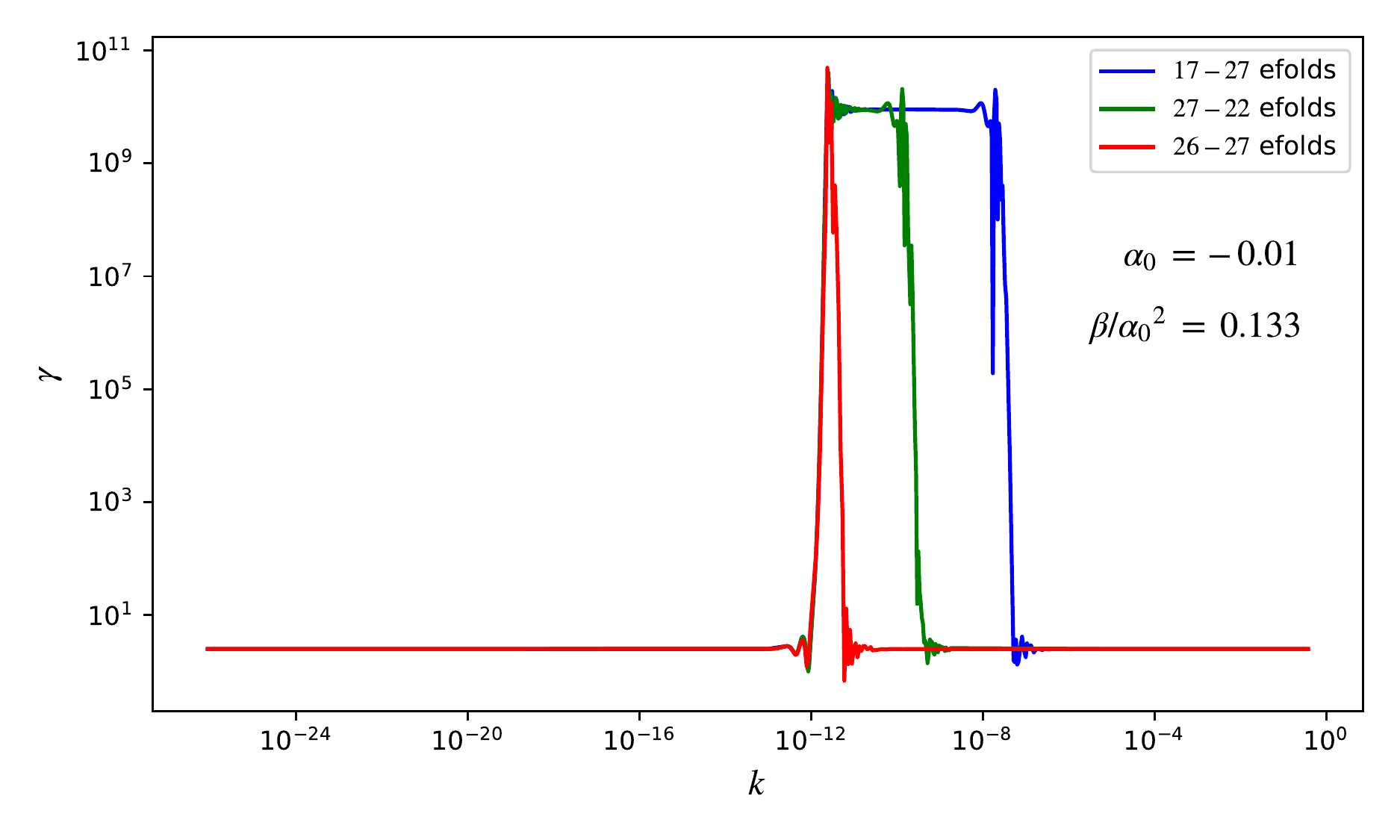}
		\includegraphics[width=0.7 \columnwidth]{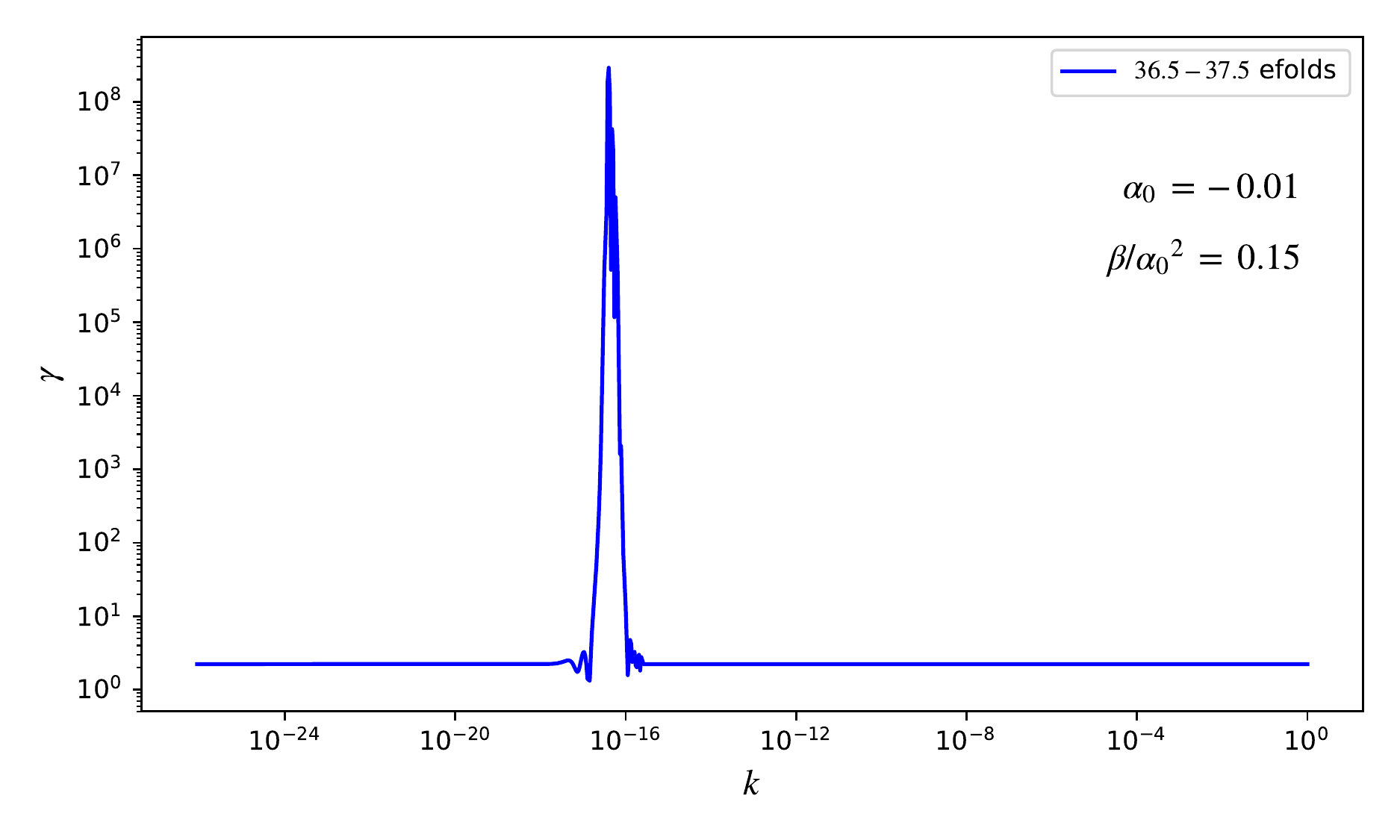}
		\caption{The upper figure illustrates the generic behavior of the modulating factor, $\gamma$. In these three different plots, we have fixed $N_1=30$ and put $N_2=20, 25, 29$ for blue, green, and red lines, respectively. The delay parameter $\delta n$ is almost the same in all these plots and turns out to be about 3 e-folds. In the lower figure, the enhancement occurs in scales $k=e^{-37.5}H-e^{-36.5}H$ would lead to the formation of PBHs with a mass about $30 M_{\odot}$.}
		\label{fig:data-behavior}}
\end{figure}
As we will review in the next section, the masses of the produced PBHs depend on the fluctuations with an amplitude of order one exiting the horizon. For instance, for the formation of PBHs of $\sim 30M_{\odot}$, the modes that exit the horizon about $23$ e-folding after the largest CMB scales, corresponding to the mode that exit $37$ e-folds before the end of inflation, should have had such an amplitude.  Fig. \ref{fig:data-behavior} shows that for $\alpha_0=0.01$ and $z=0.15$ such amplitudes for the modulation factor can be achieved.  The upper graph shows that regardless of the duration of the interval in which the quartic coupling is negative, there is a delay of almost the same magnitude, $N_d\sim 3$ e-foldings. The lower graph shows the case in which the enhancement occurs for the range of $k$, leading to the formation of PBHs of order $30 M_{\odot}$.
\subsection{EFT Compatibility}
It was pointed out in \cite{Cheung:2007st} that the pure sextic dispersion relation is not compatible with the EFT criteria, as in the low energy limit, the interacting terms in the third-order Lagrangian (terms like $\dot{\pi}(\nabla\pi)^2$), become strong at low energy. However, in the presence of quartic and quadratic terms in the dispersion relation, they are the terms that can become dominant at low energy. In particular, if at the horizon crossing, the contribution of the sixth order term in the dispersion relation is subdominant with respect to the quadratic or quartic terms, one can safely ignore the non-linear evolution inside the horizon\footnote{The interacting operator that determines the amplitude of non-gaussianity is $\dot{\pi}\nabla \pi^2$. This operator has a negative energy scaling dimension if the dispersion relation is dominant by the sixth order term in the polynomial dispersion relation. Namely the expectation value of this cubic operator grows if the sixth order term in the dispersion relation remains dominant up to horizon crossing.
 However, with the dominance of the quartic and quadratic terms in the dispersion relation, as the mode physical wavelength increases, the growth of non-gaussianity stops and it remains small enough such that the perturbativity is honored.}
  In analogy with \cite{Ashoorioon:2017toq}, such higher-order corrections to the dispersion relation is expected to only lead to $f_{\rm NL}^{\rm loc}\sim {\rm few}$, which can be made compatible with the EFT criterion,
\beq
f_{\rm NL}\times \mathcal{P}^{1/2}_{{}_S}\lesssim 1\,.
\eeq
In particular for the example we considered here with the numerical values for the coefficients $\alpha_0=-0.01$ and $\frac{\beta}{\alpha_0^2}\simeq 0.15$, one can easily check that horizon crossing occurs at $x_3\simeq -1.42883$ (this is slightly earlier than $x=-\sqrt{2}$, which happens for a Lorentzian dispersion relation $\omega^2=k^2$). This corresponds to the largest root of the effective frequency,
\beq
\omega^2(x)=c_{{}_S}^2 +\alpha x^2+\beta x^4-\frac{2}{x^2}
\eeq
with $c_{{}_S}=1$, $\alpha=-0.01$ and $\beta=0.15\alpha^2$. At $x_3$, the ratios of the sixth and fourth-order terms to the quadratic one are, respectively, $\sim 6.25\times 10^{-5}$ and $\sim 0.02$ and thus are negligible. In particular, the sixth order term, which could cause difficulty, can be safely ignored at the horizon crossing.
\section{PBH Formation}\label{PBHformation}
This section reviews the basic idea that the collapse of large densities may lead to a PBH under certain conditions. When in the radiation-dominated era, a highly over-dense region re-enters the cosmological horizon, it may overcome the pressure and collapse to find itself as a PBH. Mathematically, these over-dense regions are described by sufficiently large cosmological perturbations. Although the early universe after the inflationary era can be well-described by the FLRW metric, we need to perturb this background to understand how perturbed regions provide the opportunity of forming PBHs. Using spherical symmetry assumption, at the super-horizon scales, one can write down the following approximate form of metric to describe such regions :
\begin{equation}\label{eq1}
	ds^2 = -dt^2 + a^2(t) e^{2\psi(r)}\big[ dr^2 + r^2d\Omega^2 \big],
\end{equation}
where $a(t)$ is the scale factor and $\psi(r)$ has the role of comoving curvature perturbation. This departure from the FLRW metric leads us to a density contrast defined on comoving hyper-surface. In the gradient expansion approximation, using Einstein equations, one can show that \cite{Musco:2018rwt} the comoving curvature perturbation is nonlinearly conserved and related to the density contrast as

\begin{eqnarray}\label{eqq1}
	\frac{\delta\rho}{\rho_b} =-\frac{8}{9} \frac{1}{a^2H^2}e^{-\frac{5}{2}\psi(r)} \nabla^2 e^{\frac{1}{2}\psi(r)},
\end{eqnarray}
where $\rho_b = 3M_{P}^{2} H^{2}$ is the energy density of the background.
Initially, the comoving size of these perturbed regions are much larger than the Hubble horizon, but when the universe evolves, we expect that the comoving scales of these regions become of the order of Hubble horizon scale. Because only perturbations on scales larger than Jeans length are able to collapse to form PBHs, we need $c^2_s k^2 \approx a^2 H^2$ (with $c^2_s = \frac{1}{3}$), at the time $t=t_f$ of collapse of a over-density with wavenumber $k$ (the time of PBH formation).  This would give us a straightforward criterion which states for PBH formation the density contrast should be bigger than $c^2_s$ when the scale of interest reenters the horizon (or $k=a H$), namely
\begin{equation}\label{eq6}
	\frac{\delta\rho}{\rho}(t_k) = \frac{c^2_s k^2}{H^2 a^2}\big|_{t=t_k} \geqslant c^2_s = 1/3
\end{equation}
where $t_k$ determines the horizon crossing time for a given scale $k$. This relation also implies that  the mass of the PBH at the formation time can be approximately identified with the horizon mass. This is because one cannot here recognize a meaningful discriminant between  the Hubble horizon and Jeans length $R_J = c_{{}_S}/H$. In fact a more precise investigation \cite{Carr:1975qj} unveils that at the formation time during the radiation dominated era, the mass of PBHs is related to the horizon mass $M_H\equiv \frac{4\pi \rho_b}{3 H^3}$ as the following,
\begin{equation}\label{eq7}
	M_{{}_{\rm PBH}} = \gamma_{*} M_H\big|_{t=t_f} = \frac{\gamma_{*}}{2GH(t_f)}\,,
\end{equation}
where the correction factor $\gamma_{*}$ in the simple analytic calculation has been estimated as $\gamma_{*} \approx \frac{1}{3\sqrt{3}}$.

Because we consider the cosmological perturbations are related to a quantum origin, it would be sensible to expect that their amplitudes have a statistical distribution. In fact, in the spirit of perturbation theory, one should understand the density contrast as a statistical variable with a very small mean value  (which also implies a negligible mean value for $\psi$).   Therefore, for the PBH formation, we have to look for a large deviation from the mean value. In the case of negligible non-Gaussianities, we may consider the density contrast and the comoving curvature perturbation as approximate Gaussian variables which would be fully described by the two-point function of $\psi$ using the linearized form of Eq. (\ref{eqq1}), namely
\begin{equation}\label{eqq2}
	\frac{\delta\rho}{\rho_b} \simeq -\frac{4}{9} \frac{1}{a^2H^2}\nabla^2 \psi(r).
\end{equation}
This immediately relates the power spectrum of the comoving curvature perturbation $\mathcal{P}_{{}_S}$ with the power spectrum of the density contrast $\mathcal{P}_{\delta}$ :
\begin{equation}\label{eqq3}
	\mathcal{P}_{\delta}(k,k^{'},t) \simeq (2\pi)^3 \delta^3(k-k^{'})  \Big(\frac{k}{aH}\Big)^4 \frac{16}{81} \mathcal{P}_{{}_S}(k).
\end{equation}
Inflationary scenarios may provide us with mechanisms that using them the PBHs formation can be induced by the primordial fluctuations with this assumption that the overdense regions described above are sourced from the primordial curvature perturbations. Such mechanisms have been designed so that the amplitudes of these primordial curvature fluctuations get some large values at certain scales depended on the mass of PBHs. So we need to find out how the mass of PBHs and the comoving scale of perturbations are related to each other. Since  PBHs are formed when over-densities re-enter the horizon, we can identify the size of such regions with the comoving wavenumber  $k$ of the primordial perturbations at the formation time, i.e., $a(t_f)H(t_f) \propto k$. Because for the radiation-dominated universe, we have $H\propto \frac{1}{a^2}$, the relation between the Hubble parameter and the comoving wavenumber of primordial fluctuations turns out to be $H(t_f) \propto k^2$. Using relation (\ref{eq7}) one can deduce that $M_{{}_{\rm PBH}}(k)\propto k^{-2}$. A more rigorous analysis \cite{Kawasaki:2016pql} reveals the following approximation for scale dependency of the PBH mass
\begin{equation}\label{eq8}
	M_{{}_{\rm PBH}}(k) \approx 30\ M_{\odot}\, \big(\frac{\gamma_{*}}{0.2}\big) \big(\frac{g_{{}_{*,{\rm form}}}}{10.75} \big)^{-1/6}\Big( \frac{k}{2.9 \times 10^5 {\rm Mpc}^{-1}}\Big)^{-2}
\end{equation}
Using this relation we can estimate the scale of instabilities in the power spectrum from which  PBHs of a specific mass come out. For example, for creating
PBHs of mass $30 M_{\odot}$, the instabilities should be placed at the scale of $k \sim 2.9\times 10^5 {\rm Mpc}^{-1}$. Therefore there is a hierarchy between
the observable scales $k_{{}_{\rm CMB}} \sim 0.002 {\rm Mpc}^{-1}$ namely CMB scales, and the scales of PBHs of mass $30 M_{\odot}$ (which is not accessible by CMB) as
$k_{30M_{\odot}} \sim e^{20} k_{{}_{\rm CMB}}$. Moreover, for PBHs of the range mass $M_{{}_{\rm PBH}} \sim 10^{-17} M_{\odot}-10^{-13} M_{\odot}$ which can constitute $10\%-100\%$ dark matter abundance, the corresponding scale would be between $35.5$ and $40$ e-folds after the CMB scales exit.
\begin{figure}[t]
	\centering
	\includegraphics[width=0.7 \columnwidth]{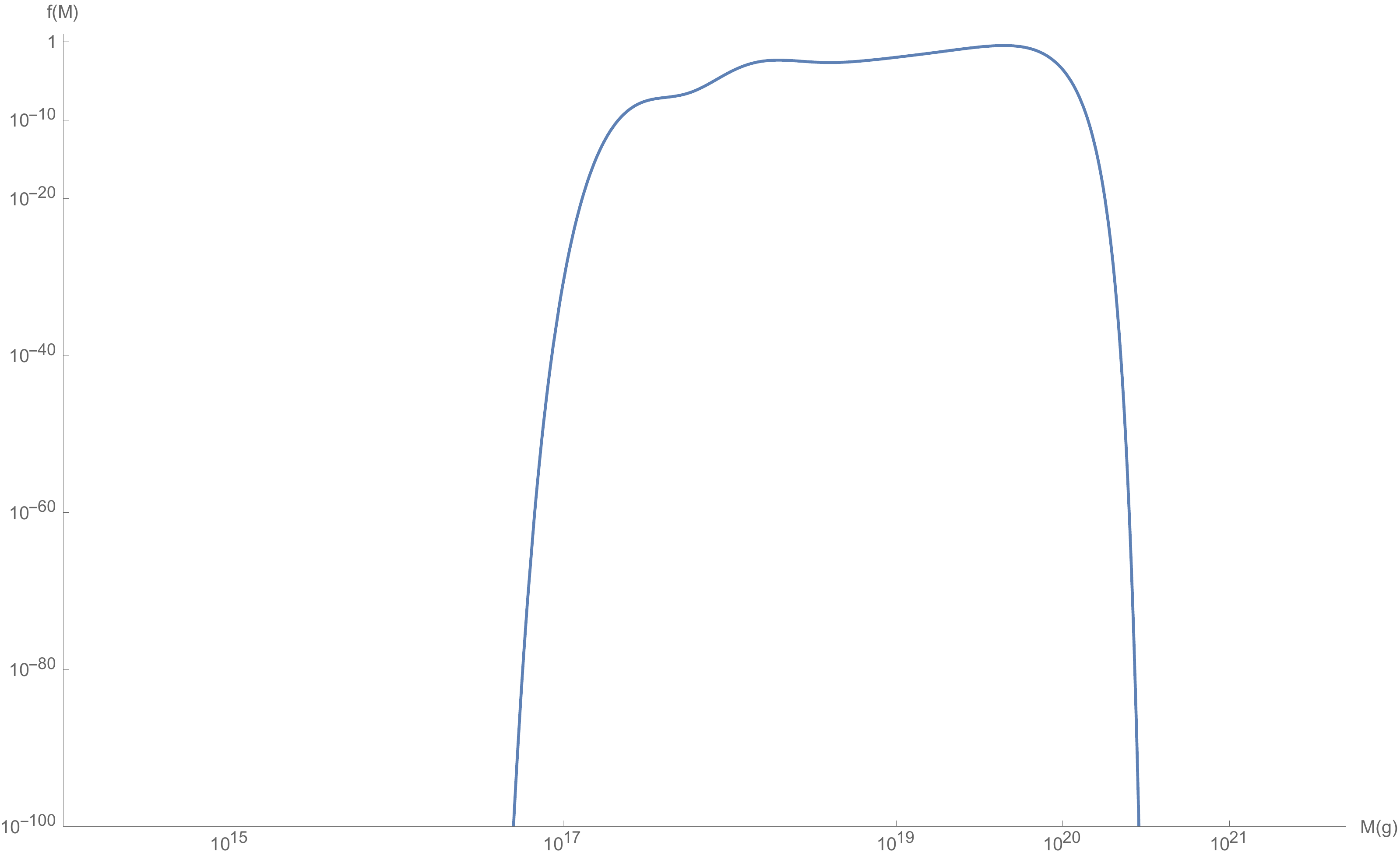}
	\caption{This figure shows the fraction of PBH mass relative to dark matter in the present time coming up from $\gamma$ in Fig. \ref{fig:alpha43}, in terms of its mass with the parameter $\delta_{th}=0.5$. }
	\label{fig:1}
\end{figure}
\section{PBHs Mass Fraction:}

In the last section, we saw how extended EFT modify the scalar perturbation power spectrum respect to the standard Bunch-Davis power spectrum. To see  the observational effect of this modification, one can investigate the abundance of formed PBHs, which
represents the mass fraction of PBHs at the formation time (radiation dominated era) as $\beta'$, which can be defined as
\beq
\beta'=\frac{\rho_{{}_{\rm PBH}}}{\rho_{cr}}|_{formation~time}
\eeq
where $\rho_{cr}$ is the total energy density. To state the condition for a collapse to a PBH is usually stated in terms of the smoothed density contrast at horizon crossing (HC), $\delta_{\rm HC}(R)$.  If $\delta_{\rm HC}(R) > \delta_{th} \sim {\cal O} (1)$~\cite{threshould1,threshould2,threshould3,threshould4} a fluctuation on a scale $R(M_{{}_{\rm PBH}})$ will collapse to form a PBH, with mass $M_{{}_{\rm PBH}}$ around the horizon mass \footnote{Though the naive calculation \cite{Carr:1975qj} tells us that the PBH formation occurs when the density perturbation becomes comparable to $\delta_{th}=1/3$, more general analysis  raises this value to $\delta_{th}=0.7$ \cite{threshould1,threshould4}.}.
In order to calculate the mass fraction, we just need to relate the PBH mass M to the comoving smoothing scale $R$ when the scale enters the horizon, $k = aH$. One can show that in the radiation dominated era
\beq
\frac{R}{1~ {\rm Mpc}}= 5.54 \times 10^{-24} \frac{1}{\gamma_{*}} \left(\frac{M_{{}_{\rm PBH}}}{1 {\rm g}}\right)^{1/2}\left(\frac{g_*}{3.36}\right)^{1/6}
\eeq

where $g_*$  the number of relativistic degrees of freedom, is expected to be of order 100 in the early universe and $\gamma_{*} \sim 0.2$ during the radiation era. To calculate  PBH mass fraction, one needs to evaluate, $\sigma(R)$,  the variance of the density fluctuations on the mass scale $M_{{}_{\rm PBH}}$
\begin{equation}
	\label{variance}
	\sigma^2(R)=\int_{0}^{\infty} \tilde{W}^2(k,R)\mathcal{P}_{{}_S}(k, t)\frac{{\rm d} k}{k}\,,
\end{equation}
for the probability distribution of the smoothed density contrast. Here $\mathcal{P}_{\delta}(k, t)$ is the power spectrum of the
(unsmoothed) density contrast, $ \mathcal{P}_{\delta}(k, t) \equiv \frac{k^3}{2 \pi^2} \langle
|\delta_{k} |^2 \rangle
$.
The $\tilde{W}(k,R)$ is the Fourier transform of the window function, which is used to smooth the density contrast. In this paper, we will use the Fourier transform of a volume-normalized Gaussian window function as
$
\tilde{W}(k,R)=\exp(-\frac{k^2 R^2}{2})\,.
$
The probability distribution function of the density fluctuations is given,
$\beta'$ can be regarded as the probability that the density contrast is larger than the threshold for PBH formation, and one can evaluate the mass fraction of PBHs for a  Gaussian distribution as
\begin{equation}
	\beta'(M_{{}_{\rm PBH}})  =
	\int_{\delta_{th}}^{\infty} P(\delta_{\rm HC}(R)) \,{\rm d} \delta_{\rm HC}(R)= \frac{1}{\sqrt{2\pi}\sigma_{\rm HC}(R)}
	\int_{\delta_{th}}^{\infty} \exp{\left(- \frac{\delta^2_{\rm HC}(R)}
		{2 \sigma_{\rm HC}^2(R)}\right)}
	\,{\rm d}\delta_{\rm HC}(R)  \,.
	\label{presssch}
\end{equation}
\begin{figure}[t]
	\centering
	\includegraphics[width=0.8 \columnwidth]{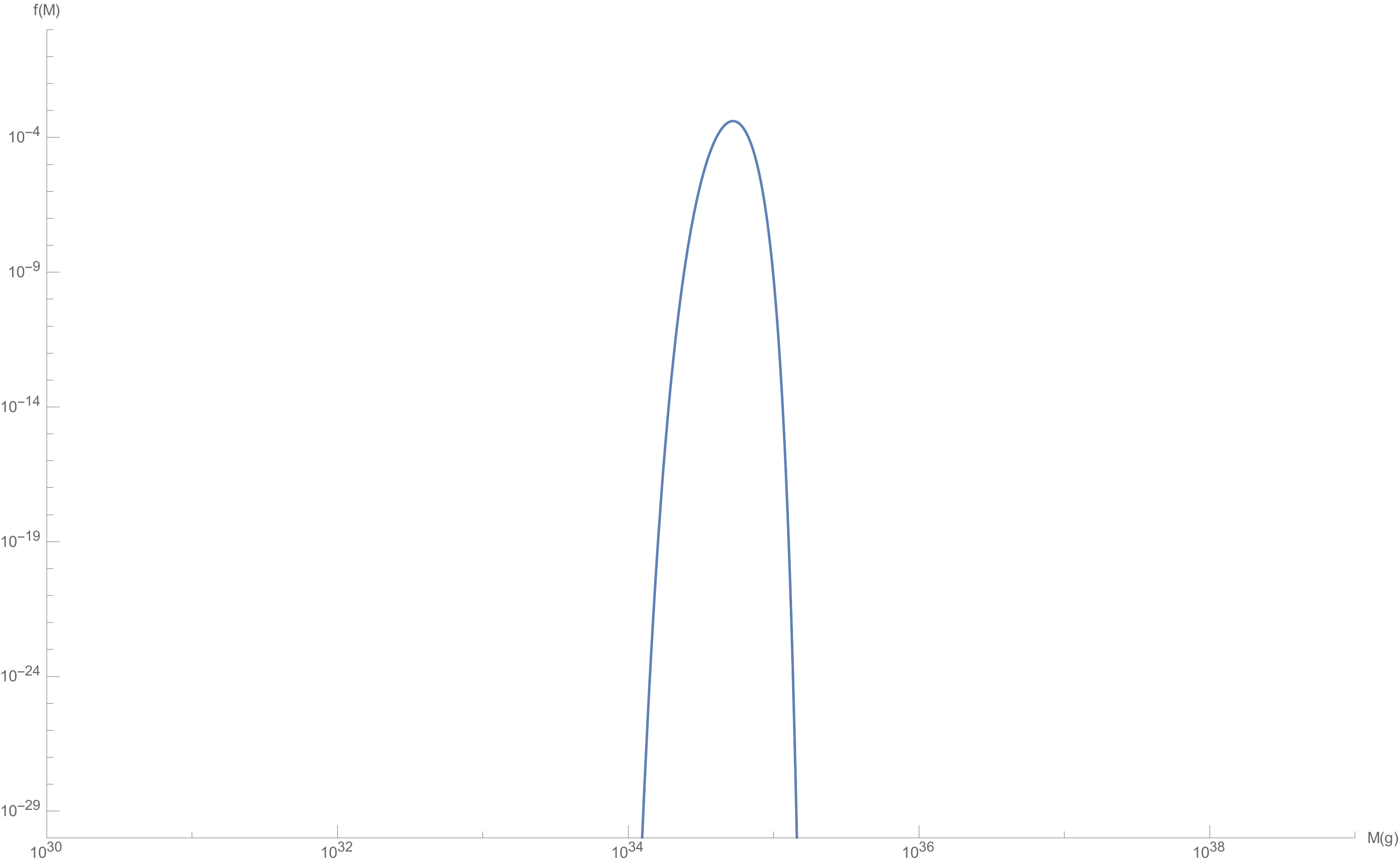}
	\caption{This figure shows the fraction of PBH mass relative to dark matter in the present time coming up from the lower blue plot in Fig.\ref{fig:data-behavior}, in terms of its mass with the parameters $\delta_{th}=0.5$.}
	\label{fig:2}
\end{figure}
Using this  mass fraction one can evaluate the fraction of PBH mass relative to dark matter in the present time as
\begin{equation}
	f(M_{{}_{\rm PBH}})=2.7 \times 10^{8}\left(\frac{\gamma_{*}}{0.2}\right)^{1/2} \left(\frac{g_{*}}{10.75}\right)^{-1/4} \left(\frac{M}{M_{\odot}}\right)^{-1/2} \beta'(M).
	\label{f}
\end{equation}
The constraints on the PBH initial mass fraction, $\beta'(M_{{}_{\rm PBH}})$, can therefore be
translated into constraints on the mass variance by inverting this expression. There is a wide range of constraints on the PBH abundance, from their various gravitational effects and the consequences of their evaporation, which apply over different mass ranges. These constraints are mass-dependent and lie in the range $\beta'(M_{{}_{\rm PBH}}) < 10^{-20} - 10^{-5}$~\cite{Josan:2009qn,Carr:2009jm,Carr:2020gox}. The power of these PBH abundance constraints is apparent when we consider the resulting constraints on $\sigma_{\rm HC}(R)$ which are in the range $\sigma_{\rm HC}(R)/\delta_{th}  < 0.1-0.2 $. In other words a small change in $\sigma_{\rm HC}(R)/\delta_{th} $ leads to a huge change in $\beta'$, and respectively $f(M)$. Figure (\ref{fig:2}) shows that we can get enough PBH which can give us a significant part of dark matter. It also demonstrates modulated oscillations in the mass function of the PBHs, which is the result of its exponential sensitivity to the changes in the power spectrum. On the other hand, figure (\ref{fig:1}) shows that we can get enough solar PBH mass which forms binaries, and we can see their merging effect in LIGO.

\section{Concluding Remarks}

In this paper, we proposed a mechanism to enhance the power spectrum at the relevant scales to the PBH formation threshold. The mechanism was based on the observation that the dispersion relation with a negative interim slope can substantially enhance the two-point function if the coefficient of the negative term in the dispersion relation is smaller than a threshold. An example of such dispersion relations is the sixth order dispersion relation embedded in the EEFToI \cite{Ashoorioon:2018uey}. In the effective field theory approach, even if the coefficients of the operators in the unitary gauge action are time-dependent, the spatial diffeomorphism is respected. Since the coefficients of the sixth order dispersion relation are related to these time-dependent coefficients, we conceived a situation that the quartic coupling swings from a small positive value to a small negative one, which lasts for some number of e-folds, and then becomes positive again, while the sextic and the speed of sound were kept nearly constant. We also confined ourselves to the parameter space region in which the sound speed remained close to one. This way, we would be able to avoid large non-gaussianities. We assumed that these changes of signs happen almost instantaneously, and thus we can safely ignore the transient effects. Interestingly, we found the impact of this change of sign in the dispersion relation is imprinted on the power spectrum of perturbations only after a couple of e-folds, confirming that the enhancement occurs while the modes are well inside the horizon. The amplification occurs because of the influence of the negative contribution of the quartic term in the dispersion relation, which leads to an exponential growth in the mode function, subsequently the power spectrum. The smaller the absolute value of the quartic coefficient, the longer the mode spends under the influence of this term, hence the larger the amplitude of the power spectrum. Also, if the absolute value of the coefficient of the quartic term in the dispersion over the sixth order one is smaller than a threshold, this enhancement will be more pronounced. We showed that for reasonable values of the quartic and the sixth order coefficients in the dispersion relation, the power spectrum is augmented with the desired modulation factor, although the sixth order term in the dispersion relation is negligible in comparison with the quartic one which itself, in turn, is small in comparison with the Lorentzian term, $k^2$. This helps us to avoid the strong coupling at the horizon crossing.

As for higher order terms, as far as their coefficient is positive and small, there will not be that much of modulation in the amplitude of the power spectrum. In fact, higher order positive terms tend to suppress the amplitude of scalar density perturbations (although very mildly). Only when the coefficient of one of these higher order terms becomes negative, there is a possibility that one can enhance the power spectrum. Although in this article, we designed our scenario with a the sixth order polynomial dispersion relation, in principle this could have also been done with higher order polynomials. The only requirements are that one is able to define a stable vacuum for the dispersion relation when the mode is deep inside the horizon and also there is an intermediate term which appears with a negative sign in the dispersion relation. If the negative intermediate term lasts long enough, the desired enhancement of the power spectrum can be achieved. The reason we focused on the sextic polynomial dispersion relation, is that we have an explicit construction of the polynomial dispersion relation in the context of Extended EFT of inflation in \cite{Ashoorioon:2018uey}.

Not only it can be arranged that the Extended EFT of inflation, with parameters chosen so as to it leads to PBH formation, avoids the strong coupling in the IR and, in particular, the horizon crossing, but it can also be arranged such that the new UV cutoff of the theory remains well above the Hubble parameter. As pointed out in \cite{Ashoorioon:2018uey}, the UV cutoff in the Extended EFToI also comes from the same four-leg interacting operator $(\partial^i \pi \partial_i \pi)^2$, which has the energy scaling dimension of $2/3$ at high energies. The coefficient of this dispersion relation is given by $M_2^4$, which is the coefficient of $(1+g^{00})^2$ term in the unitary gauge action. On the other hand, the canonical field $\pi_c\sim \sqrt{A_0}\pi $, where $A_0$ is defined as,
\begin{equation}
	A_0=G_1+ 9 H^4 (3\delta_3+\frac{3}{2}\delta_4)-9\bar{M}_4 H^3\,,
\end{equation}
where $G_1$ is defined in eq. \eqref{var2}. The cutoff of the theory then turns out to be
\begin{equation}
	\Lambda^4 \simeq \frac{c_{{}_S}^7 A_0^2}{M_2^4}\,,
\end{equation}
where now $c_{{}_S}$ is given contains all the parameters of the Extended EFT of inflation. In principle, by adjusting the coefficients in the unitary gauge action, one can set the cutoff large enough such that $\Lambda\gg H$. In this work, we have assumed that the coefficients and couplings in unitary gauge action were chosen such that $c_{{}_S}\approx 1$, Now if  $A_0^2$ is arranged to be much bigger than $(4 M_2^4-2 M_{\rm Pl}^2 \dot{H})^2$ and $M_2^4 H^4$, then the UV cutoff in the Extended EFT of inflation would not only be much bigger than the Hubble parameter during inflation but also UV cutoff in the EFToI \cite{Cheung:2007st}.

In the above scenario, we had assumed that the quartic coefficient becomes negative around the scale of interest and then swings back to positive values at smaller scales. This was to avoid the overproduction of PBHs in the mass ranges that are not allowed by observation. However, this may be avoided if one takes into account the fact that the thermal history of the universe and sudden drops in the pressure of relativistic matter at $W^{\pm}/Z^0$ decoupling, the quark--hadron transition and $e^+ e^-$ annihilation increases the probability of primordial black hole (PBH) formation in the early universe and implies several peaks in the  PBH mass spectrum at $10^{-6}$, 1, 30, and $10^6 ~M_{\odot}$ \cite{Carr:2019kxo}. In this scenario, it is still required that the primordial power spectrum is enhanced to order ${\rm few} \times 0.01$.

Due to the smallness of the sextic and quartic terms at the moment of horizon crossing, and as we have focused on the region of parameter space in which the speed of propagations of fluctuation is luminal, we expect that the non-gaussianity resulted from the setup to be small. However, the resulted non-gaussianity from the setup still can be larger than the slow-roll inflationary background. For the case of dispersion relations with the quartic correction to the dispersion relation, predicted in the EFT of inflation \cite{Cheung:2007st}, considering interacting operators, it has been shown that the flattened configuration, in which two of the momenta is collinear with the third one $k_1=k_2+k_3$, could be enhanced. This may compensate for the small slow-roll suppressed non-gaussianity and bring $f_{{}_{\rm NL}}^{{}^{\rm flat}}$ to the order of {\it few}. With a finite value for the non-gaussianity, it might be possible to produce PBHs from the tail of the non-gaussianity solely, as in \cite{Atal:2019erb} or \cite{Panagopoulos:2019ail}. This gives us further motivation to look at the amount of generated non-gaussianity more closely in this scenario in a separate work.

Last but not least, it is worth mentioning that the kind of dispersion relation that we elaborated on seems to be difficult to be realized within a
fundamental theories, such as ``local quantum field theory'' or perturbative string theory. As it has been pointed out by \cite{Adams:2006sv}, higher order
dispersion relations can only be implemented in such fundamental theories, if the coefficients of the higher order derivative interaction terms appear with
positive values in the Lagrangian. This corresponds to the dispersion relations in which the coefficient of higher order corrections to the Lorentzian
dispersion relation, $\omega^2=k^2$, appear with negative sign. Otherwise, one will have superluminal propagations. This is despite the fact that the theory's
Lagrangian may looks like a local Lorentz-invariant one. This is true for both our Extended EFT of inflation, with the sextic polynomial dispersion relation, and the ghost
inflation with quartic (polynomial) dispersion relation. In both cases one can write down an ``apparently-local'' form for the Lagrangian. However, looking at
the perturbations, one can see  that the speed of propagation is larger than the speed of light. Although in part of the parameter space that we focus on
the quartic contribution to the dispersion relation appears with negative sign, the sixth order term has a positive coefficient, which gives rise to
superluminal propagation,  in particular in the regime when the mode has a very small wavelength. It is conceivable that before the dispersion relation becomes
superluminal under the influence of the sixth order term, the Lorentzian dispersion relation is recovered at very small wavelengths.

	%
	\bibliographystyle{JHEP}
	\bibliography{bibtex}

\providecommand{\href}[2]{#2}\begingroup\raggedright\begin{thebibliography}{10}

\bibitem{TheLIGOScientific:2016src}
{\bf LIGO Scientific, Virgo} Collaboration, B.~P. Abbott {\em et.~al.}, {\it
  {Tests of general relativity with GW150914}},  {\em Phys. Rev. Lett.} {\bf
  116} (2016), no.~22 221101, [\href{http://arXiv.org/abs/1602.03841}{{\tt
  1602.03841}}]. [Erratum: Phys. Rev. Lett.121,no.12,129902(2018)].

\bibitem{Abbott:2016nmj}
{\bf LIGO Scientific, Virgo} Collaboration, B.~P. Abbott {\em et.~al.}, {\it
  {GW151226: Observation of Gravitational Waves from a 22-Solar-Mass Binary
  Black Hole Coalescence}},  {\em Phys. Rev. Lett.} {\bf 116} (2016), no.~24
  241103, [\href{http://arXiv.org/abs/1606.04855}{{\tt 1606.04855}}].

\bibitem{Bird:2016dcv}
S.~Bird, I.~Cholis, J.~B. Munoz, Y.~Ali-Haimoud, M.~Kamionkowski, E.~D. Kovetz,
  A.~Raccanelli, and A.~G. Riess, {\it {Did LIGO detect dark matter?}},  {\em
  Phys. Rev. Lett.} {\bf 116} (2016), no.~20 201301,
  [\href{http://arXiv.org/abs/1603.00464}{{\tt 1603.00464}}].

\bibitem{Sasaki:2016jop}
M.~Sasaki, T.~Suyama, T.~Tanaka, and S.~Yokoyama, {\it {Primordial Black Hole
  Scenario for the Gravitational-Wave Event GW150914}},  {\em Phys. Rev. Lett.}
  {\bf 117} (2016), no.~6 061101, [\href{http://arXiv.org/abs/1603.08338}{{\tt
  1603.08338}}]. [erratum: Phys. Rev. Lett.121,no.5,059901(2018)].

\bibitem{Clesse:2016vqa}
S.~Clesse and J.~Garcia-Bellido, {\it {The clustering of massive Primordial
  Black Holes as Dark Matter: measuring their mass distribution with Advanced
  LIGO}},  {\em Phys. Dark Univ.} {\bf 15} (2017) 142--147,
  [\href{http://arXiv.org/abs/1603.05234}{{\tt 1603.05234}}].

\bibitem{Hawking:1982ga}
S.~W. Hawking, I.~G. Moss, and J.~M. Stewart, {\it {Bubble Collisions in the
  Very Early Universe}},  {\em Phys. Rev.} {\bf D26} (1982) 2681.

\bibitem{Crawford:1982yz}
M.~Crawford and D.~N. Schramm, {\it {Spontaneous Generation of Density
  Perturbations in the Early Universe}},  {\em Nature} {\bf 298} (1982)
  538--540. [,333(1982)].

\bibitem{Deng:2017uwc}
H.~Deng and A.~Vilenkin, {\it {Primordial black hole formation by vacuum
  bubbles}},  {\em JCAP} {\bf 1712} (2017), no.~12 044,
  [\href{http://arXiv.org/abs/1710.02865}{{\tt 1710.02865}}].

\bibitem{Deng:2016vzb}
H.~Deng, J.~Garriga, and A.~Vilenkin, {\it {Primordial black hole and wormhole
  formation by domain walls}},  {\em JCAP} {\bf 1704} (2017), no.~04 050,
  [\href{http://arXiv.org/abs/1612.03753}{{\tt 1612.03753}}].

\bibitem{Ashoorioon:2020hln}
A.~Ashoorioon, A.~Rostami, and J.~T. Firouzjaee, {\it {Charting the Landscape
  in Our Neighborhood from the PBHs Mass Distribution and GWs}},
  \href{http://arXiv.org/abs/2012.02817}{{\tt 2012.02817}}.

\bibitem{Hawking:1987bn}
S.~W. Hawking, {\it {Black Holes From Cosmic Strings}},  {\em Phys. Lett.} {\bf
  B231} (1989) 237--239.

\bibitem{Polnarev:1988dh}
A.~Polnarev and R.~Zembowicz, {\it {Formation of Primordial Black Holes by
  Cosmic Strings}},  {\em Phys. Rev.} {\bf D43} (1991) 1106--1109.

\bibitem{Sanidas:2012ee}
S.~A. Sanidas, R.~A. Battye, and B.~W. Stappers, {\it {Constraints on cosmic
  string tension imposed by the limit on the stochastic gravitational wave
  background from the European Pulsar Timing Array}},  {\em Phys. Rev.} {\bf
  D85} (2012) 122003, [\href{http://arXiv.org/abs/1201.2419}{{\tt 1201.2419}}].

\bibitem{Ringeval:2017eww}
C.~Ringeval and T.~Suyama, {\it {Stochastic gravitational waves from cosmic
  string loops in scaling}},  {\em JCAP} {\bf 12} (2017) 027,
  [\href{http://arXiv.org/abs/1709.03845}{{\tt 1709.03845}}].

\bibitem{Ade:2013xla}
{\bf Planck} Collaboration, P.~A.~R. Ade {\em et.~al.}, {\it {Planck 2013
  results. XXV. Searches for cosmic strings and other topological defects}},
  {\em Astron. Astrophys.} {\bf 571} (2014) A25,
  [\href{http://arXiv.org/abs/1303.5085}{{\tt 1303.5085}}].

\bibitem{Jenkins:2020ctp}
A.~C. Jenkins and M.~Sakellariadou, {\it {Primordial black holes from cusp
  collapse on cosmic strings}},  \href{http://arXiv.org/abs/2006.16249}{{\tt
  2006.16249}}.

\bibitem{Khlopov:1980mg}
M.~Khlopov and A.~Polnarev, {\it {Primordial Black Holes as a Cosmological Test
  of Grand Unification}},  {\em Phys. Lett. B} {\bf 97} (1980) 383--387.

\bibitem{Carr:1974nx}
B.~J. Carr and S.~W. Hawking, {\it {Black holes in the early Universe}},  {\em
  Mon. Not. Roy. Astron. Soc.} {\bf 168} (1974) 399--415.

\bibitem{Schutz:2016khr}
K.~Schutz and A.~Liu, {\it {Pulsar timing can constrain primordial black holes
  in the LIGO mass window}},  {\em Phys. Rev.} {\bf D95} (2017), no.~2 023002,
  [\href{http://arXiv.org/abs/1610.04234}{{\tt 1610.04234}}].

\bibitem{Monroy-Rodriguez:2014ula}
M.~A. Monroy-Rodriguez and C.~Allen, {\it {The end of the MACHO era- revisited:
  new limits on MACHO masses from halo wide binaries}},  {\em Astrophys. J.}
  {\bf 790} (2014), no.~2 159, [\href{http://arXiv.org/abs/1406.5169}{{\tt
  1406.5169}}].

\bibitem{Gaggero:2016dpq}
D.~Gaggero, G.~Bertone, F.~Calore, R.~M.~T. Connors, M.~Lovell, S.~Markoff, and
  E.~Storm, {\it {Searching for Primordial Black Holes in the radio and X-ray
  sky}},  {\em Phys. Rev. Lett.} {\bf 118} (2017), no.~24 241101,
  [\href{http://arXiv.org/abs/1612.00457}{{\tt 1612.00457}}].

\bibitem{Yokoyama:1998xd}
J.~Yokoyama, {\it {Cosmological constraints on primordial black holes produced
  in the near critical gravitational collapse}},  {\em Phys. Rev.} {\bf D58}
  (1998) 107502, [\href{http://arXiv.org/abs/gr-qc/9804041}{{\tt
  gr-qc/9804041}}].

\bibitem{Carr:2020gox}
B.~Carr, K.~Kohri, Y.~Sendouda, and J.~Yokoyama, {\it {Constraints on
  Primordial Black Holes}},  \href{http://arXiv.org/abs/2002.12778}{{\tt
  2002.12778}}.

\bibitem{Akrami:2018odb}
{\bf Planck} Collaboration, Y.~Akrami {\em et.~al.}, {\it {Planck 2018 results.
  X. Constraints on inflation}},  \href{http://arXiv.org/abs/1807.06211}{{\tt
  1807.06211}}.

\bibitem{Carr:1993aq}
B.~J. Carr and J.~E. Lidsey, {\it {Primordial black holes and generalized
  constraints on chaotic inflation}},  {\em Phys. Rev.} {\bf D48} (1993)
  543--553.

\bibitem{Josan:2009qn}
A.~S. Josan, A.~M. Green, and K.~A. Malik, {\it {Generalised constraints on the
  curvature perturbation from primordial black holes}},  {\em Phys. Rev.} {\bf
  D79} (2009) 103520, [\href{http://arXiv.org/abs/0903.3184}{{\tt 0903.3184}}].

\bibitem{Cole:2017gle}
P.~S. Cole and C.~T. Byrnes, {\it {Extreme scenarios: the tightest possible
  constraints on the power spectrum due to primordial black holes}},  {\em
  JCAP} {\bf 02} (2018) 019, [\href{http://arXiv.org/abs/1706.10288}{{\tt
  1706.10288}}].

\bibitem{Germani:2018jgr}
C.~Germani and I.~Musco, {\it {Abundance of Primordial Black Holes Depends on
  the Shape of the Inflationary Power Spectrum}},  {\em Phys. Rev. Lett.} {\bf
  122} (2019), no.~14 141302, [\href{http://arXiv.org/abs/1805.04087}{{\tt
  1805.04087}}].

\bibitem{Green:1997sz}
A.~M. Green and A.~R. Liddle, {\it {Constraints on the density perturbation
  spectrum from primordial black holes}},  {\em Phys. Rev.} {\bf D56} (1997)
  6166--6174, [\href{http://arXiv.org/abs/astro-ph/9704251}{{\tt
  astro-ph/9704251}}].

\bibitem{Kim:1996hr}
H.~I. Kim and C.~H. Lee, {\it {Constraints on the spectral index from
  primordial black holes}},  {\em Phys. Rev.} {\bf D54} (1996) 6001--6007.

\bibitem{Stewart:1997wg}
E.~D. Stewart, {\it {Flattening the inflaton's potential with quantum
  corrections. 2.}},  {\em Phys. Rev.} {\bf D56} (1997) 2019--2023,
  [\href{http://arXiv.org/abs/hep-ph/9703232}{{\tt hep-ph/9703232}}].

\bibitem{Drees:2011hb}
M.~Drees and E.~Erfani, {\it {Running-Mass Inflation Model and Primordial Black
  Holes}},  {\em JCAP} {\bf 1104} (2011) 005,
  [\href{http://arXiv.org/abs/1102.2340}{{\tt 1102.2340}}].

\bibitem{Drees:2011yz}
M.~Drees and E.~Erfani, {\it {Running Spectral Index and Formation of
  Primordial Black Hole in Single Field Inflation Models}},  {\em JCAP} {\bf
  1201} (2012) 035, [\href{http://arXiv.org/abs/1110.6052}{{\tt 1110.6052}}].

\bibitem{Carr:2016drx}
B.~Carr, F.~Kuhnel, and M.~Sandstad, {\it {Primordial Black Holes as Dark
  Matter}},  {\em Phys. Rev.} {\bf D94} (2016), no.~8 083504,
  [\href{http://arXiv.org/abs/1607.06077}{{\tt 1607.06077}}].

\bibitem{Sasaki:2018dmp}
M.~Sasaki, T.~Suyama, T.~Tanaka, and S.~Yokoyama, {\it {Primordial black
  holes--perspectives in gravitational wave astronomy}},  {\em Class. Quant.
  Grav.} {\bf 35} (2018), no.~6 063001,
  [\href{http://arXiv.org/abs/1801.05235}{{\tt 1801.05235}}].

\bibitem{Allahverdi:2006iq}
R.~Allahverdi, K.~Enqvist, J.~Garcia-Bellido, and A.~Mazumdar, {\it {Gauge
  invariant MSSM inflaton}},  {\em Phys. Rev. Lett.} {\bf 97} (2006) 191304,
  [\href{http://arXiv.org/abs/hep-ph/0605035}{{\tt hep-ph/0605035}}].

\bibitem{Garcia-Bellido:2017mdw}
J.~Garcia-Bellido and E.~Ruiz~Morales, {\it {Primordial black holes from single
  field models of inflation}},  {\em Phys. Dark Univ.} {\bf 18} (2017) 47--54,
  [\href{http://arXiv.org/abs/1702.03901}{{\tt 1702.03901}}].

\bibitem{Germani:2017bcs}
C.~Germani and T.~Prokopec, {\it {On primordial black holes from an inflection
  point}},  {\em Phys. Dark Univ.} {\bf 18} (2017) 6--10,
  [\href{http://arXiv.org/abs/1706.04226}{{\tt 1706.04226}}].

\bibitem{Motohashi:2017kbs}
H.~Motohashi and W.~Hu, {\it {Primordial Black Holes and Slow-Roll Violation}},
   {\em Phys. Rev.} {\bf D96} (2017), no.~6 063503,
  [\href{http://arXiv.org/abs/1706.06784}{{\tt 1706.06784}}].

\bibitem{Cicoli:2018asa}
M.~Cicoli, V.~A. Diaz, and F.~G. Pedro, {\it {Primordial Black Holes from
  String Inflation}},  {\em JCAP} {\bf 1806} (2018), no.~06 034,
  [\href{http://arXiv.org/abs/1803.02837}{{\tt 1803.02837}}].

\bibitem{Clesse:2015wea}
S.~Clesse and J.~Garcia-Bellido, {\it {Massive Primordial Black Holes from
  Hybrid Inflation as Dark Matter and the seeds of Galaxies}},  {\em Phys.
  Rev.} {\bf D92} (2015), no.~2 023524,
  [\href{http://arXiv.org/abs/1501.07565}{{\tt 1501.07565}}].

\bibitem{Linde:1993cn}
A.~D. Linde, {\it {Hybrid inflation}},  {\em Phys. Rev.} {\bf D49} (1994)
  748--754, [\href{http://arXiv.org/abs/astro-ph/9307002}{{\tt
  astro-ph/9307002}}].

\bibitem{Kawasaki:2006zv}
M.~Kawasaki, T.~Takayama, M.~Yamaguchi, and J.~Yokoyama, {\it {Power Spectrum
  of the Density Perturbations From Smooth Hybrid New Inflation Model}},  {\em
  Phys. Rev.} {\bf D74} (2006) 043525,
  [\href{http://arXiv.org/abs/hep-ph/0605271}{{\tt hep-ph/0605271}}].

\bibitem{Cai:2019bmk}
R.-G. Cai, Z.-K. Guo, J.~Liu, L.~Liu, and X.-Y. Yang, {\it {Primordial black
  holes and gravitational waves from parametric amplification of curvature
  perturbations}},  {\em JCAP} {\bf 06} (2020) 013,
  [\href{http://arXiv.org/abs/1912.10437}{{\tt 1912.10437}}].

\bibitem{Ozsoy:2018flq}
O.~Ozsoy, S.~Parameswaran, G.~Tasinato, and I.~Zavala, {\it {Mechanisms for
  Primordial Black Hole Production in String Theory}},  {\em JCAP} {\bf 1807}
  (2018) 005, [\href{http://arXiv.org/abs/1803.07626}{{\tt 1803.07626}}].

\bibitem{ArmendarizPicon:1999rj}
C.~Armendariz-Picon, T.~Damour, and V.~F. Mukhanov, {\it {k - inflation}},
  {\em Phys. Lett.} {\bf B458} (1999) 209--218,
  [\href{http://arXiv.org/abs/hep-th/9904075}{{\tt hep-th/9904075}}].

\bibitem{Alishahiha:2004eh}
M.~Alishahiha, E.~Silverstein, and D.~Tong, {\it {DBI in the sky}},  {\em Phys.
  Rev.} {\bf D70} (2004) 123505,
  [\href{http://arXiv.org/abs/hep-th/0404084}{{\tt hep-th/0404084}}].

\bibitem{Cheung:2007st}
C.~Cheung, P.~Creminelli, A.~L. Fitzpatrick, J.~Kaplan, and L.~Senatore, {\it
  {The Effective Field Theory of Inflation}},  {\em JHEP} {\bf 03} (2008) 014,
  [\href{http://arXiv.org/abs/0709.0293}{{\tt 0709.0293}}].

\bibitem{Martin:2000xs}
J.~Martin and R.~H. Brandenberger, {\it {The TransPlanckian problem of
  inflationary cosmology}},  {\em Phys. Rev.} {\bf D63} (2001) 123501,
  [\href{http://arXiv.org/abs/hep-th/0005209}{{\tt hep-th/0005209}}].

\bibitem{Ashoorioon:2011eg}
A.~Ashoorioon, D.~Chialva, and U.~Danielsson, {\it {Effects of Nonlinear
  Dispersion Relations on Non-Gaussianities}},  {\em JCAP} {\bf 1106} (2011)
  034, [\href{http://arXiv.org/abs/1104.2338}{{\tt 1104.2338}}].

\bibitem{Ashoorioon:2018uey}
A.~Ashoorioon, R.~Casadio, M.~Cicoli, G.~Geshnizjani, and H.~J. Kim, {\it
  {Extended Effective Field Theory of Inflation}},  {\em JHEP} {\bf 02} (2018)
  172, [\href{http://arXiv.org/abs/1802.03040}{{\tt 1802.03040}}].

\bibitem{Weinberg:2008hq}
S.~Weinberg, {\it {Effective Field Theory for Inflation}},  {\em Phys. Rev.}
  {\bf D77} (2008) 123541, [\href{http://arXiv.org/abs/0804.4291}{{\tt
  0804.4291}}].

\bibitem{Ashoorioon:2018ocr}
A.~Ashoorioon, {\it {Non-Unitary Evolution in the General Extended EFT of
  Inflation \& Excited Initial States}},  {\em JHEP} {\bf 12} (2018) 012,
  [\href{http://arXiv.org/abs/1807.06511}{{\tt 1807.06511}}].

\bibitem{Ostrogradsky:1850fid}
M.~Ostrogradsky, {\it {Memoires sur les equations differentielles, relatives au
  probleme des isoperimetres}},  {\em Mem. Acad. St. Petersbourg} {\bf 6}
  (1850), no.~4 385--517.

\bibitem{Ashoorioon:2017toq}
A.~Ashoorioon, R.~Casadio, G.~Geshnizjani, and H.~J. Kim, {\it {Getting
  Super-Excited with Modified Dispersion Relations}},  {\em JCAP} {\bf 1709}
  (2017), no.~09 008, [\href{http://arXiv.org/abs/1702.06101}{{\tt
  1702.06101}}].

\bibitem{Musco:2018rwt}
I.~Musco, {\it {The threshold for primordial black holes: dependence on the
  shape of the cosmological perturbations}},  {\em Phys. Rev.} {\bf D100}
  (2019), no.~12 123524, [\href{http://arXiv.org/abs/1809.02127}{{\tt
  1809.02127}}].

\bibitem{Carr:1975qj}
B.~J. Carr, {\it {The Primordial black hole mass spectrum}},  {\em Astrophys.
  J.} {\bf 201} (1975) 1--19.

\bibitem{Kawasaki:2016pql}
M.~Kawasaki, A.~Kusenko, Y.~Tada, and T.~T. Yanagida, {\it {Primordial black
  holes as dark matter in supergravity inflation models}},  {\em Phys. Rev.}
  {\bf D94} (2016), no.~8 083523, [\href{http://arXiv.org/abs/1606.07631}{{\tt
  1606.07631}}].

\bibitem{threshould1}
J.~C. Niemeyer and K.~Jedamzik, {\it {Dynamics of primordial black hole
  formation}},  {\em Phys. Rev.} {\bf D59} (1999) 124013,
  [\href{http://arXiv.org/abs/astro-ph/9901292}{{\tt astro-ph/9901292}}].

\bibitem{threshould2}
M.~Shibata and M.~Sasaki, {\it {Black hole formation in the Friedmann universe:
  Formulation and computation in numerical relativity}},  {\em Phys. Rev.} {\bf
  D60} (1999) 084002, [\href{http://arXiv.org/abs/gr-qc/9905064}{{\tt
  gr-qc/9905064}}].

\bibitem{threshould3}
A.~G. Polnarev and I.~Musco, {\it {Curvature profiles as initial conditions for
  primordial black hole formation}},  {\em Class. Quant. Grav.} {\bf 24} (2007)
  1405--1432, [\href{http://arXiv.org/abs/gr-qc/0605122}{{\tt gr-qc/0605122}}].

\bibitem{threshould4}
A.~Allahyari, J.~T. Firouzjaee, and A.~A. Abolhasani, {\it {Primordial black
  holes in linear and non-linear regimes}},  {\em JCAP} {\bf 1706} (2017),
  no.~06 041, [\href{http://arXiv.org/abs/1608.00591}{{\tt 1608.00591}}].

\bibitem{Carr:2009jm}
B.~J. Carr, K.~Kohri, Y.~Sendouda, and J.~Yokoyama, {\it {New cosmological
  constraints on primordial black holes}},  {\em Phys. Rev.} {\bf D81} (2010)
  104019, [\href{http://arXiv.org/abs/0912.5297}{{\tt 0912.5297}}].

\bibitem{Carr:2019kxo}
B.~Carr, S.~Clesse, J.~Garcia-Bellido, and F.~Kuhnel, {\it {Cosmic Conundra
  Explained by Thermal History and Primordial Black Holes}},  {\em Phys. Dark
  Univ.} {\bf 31} (2021) 100755, [\href{http://arXiv.org/abs/1906.08217}{{\tt
  1906.08217}}].

\bibitem{Atal:2019erb}
V.~Atal, J.~Cid, A.~Escriva, and J.~Garriga, {\it {PBH in single field
  inflation: the effect of shape dispersion and non-Gaussianities}},
  \href{http://arXiv.org/abs/1908.11357}{{\tt 1908.11357}}.

\bibitem{Panagopoulos:2019ail}
G.~Panagopoulos and E.~Silverstein, {\it {Primordial Black Holes from
  non-Gaussian tails}},  \href{http://arXiv.org/abs/1906.02827}{{\tt
  1906.02827}}.

\bibitem{Adams:2006sv}
A.~Adams, N.~Arkani-Hamed, S.~Dubovsky, A.~Nicolis, and R.~Rattazzi, {\it
  {Causality, analyticity and an IR obstruction to UV completion}},  {\em JHEP}
  {\bf 10} (2006) 014, [\href{http://arXiv.org/abs/hep-th/0602178}{{\tt
  hep-th/0602178}}].

\end{thebibliography}\endgroup
\end{document}